\RequirePackage{fixltx2e}
\documentclass[aip,jcp,reprint,showpacs,singlecolumn]{revtex4-1}

\usepackage{dcolumn}
\usepackage{placeins}
\usepackage{epsf}
\usepackage{subfigure}
\usepackage{epsfig,amsopn}
\usepackage{amssymb,graphicx}
\usepackage{amsmath,amsfonts}
\usepackage{graphicx}
\usepackage{dcolumn}
\usepackage{bm}
\usepackage{ifpdf}
\usepackage{natbib}
\usepackage{wasysym}
\usepackage[byname]{smartref}
\usepackage{color}
\usepackage{lipsum}
\usepackage{nicefrac}
\usepackage[breaklinks, colorlinks, linkcolor=blue, citecolor=blue, urlcolor=blue]{hyperref}
\usepackage[table]{xcolor}
\usepackage{multirow}
\usepackage{braket}
\usepackage{float}
\usepackage{comment}
\usepackage{enumerate}
\usepackage{enumerate}
\usepackage{tabularx}

\newcommand{\beq}{\begin{equation}}
\newcommand{\eeq}{\end{equation}}

\newcommand{\bea}{\begin{eqnarray}}
\newcommand{\eea}{\end{eqnarray}}

\begin{document}

\title{Intermediate coherent-incoherent charge transport: DNA as a case study}

\author{Hyehwang Kim}
\affiliation{Chemical Physics Theory Group, Department of Chemistry, University of Toronto,
80 St. George Street Toronto, Ontario, Canada M5S 3H6}

\author{Michael Kilgour}
%\email{michael.kilgour@mail.utoronto.ca}
\affiliation{Chemical Physics Theory Group, Department of Chemistry, University of Toronto,
80 St. George Street Toronto, Ontario, Canada M5S 3H6}

\author{Dvira Segal}
\email{dsegal@chem.utoronto.ca}
\affiliation{Chemical Physics Theory Group, Department of Chemistry, University of Toronto,
80 St. George Street Toronto, Ontario, Canada M5S 3H6}

%\title{Intermediate coherent-incoherent charge transport: DNA as a case study}

%%%\begin{document}

\date{\today}
\begin{abstract}
We study an intermediate quantum coherent-incoherent charge transport mechanism in metal-molecule-metal junctions using B\"uttiker's probe technique.
This tool allows us to include incoherent effects in a controlled manner, and thus to study situations
in which partial decoherence affects charge transfer dynamics.
Motivated by recent experiments on intermediate coherent-incoherent charge conduction
in DNA molecules [L. Xiang {\it et al.}, Nature Chem. 7, 221-226 (2015)],
we focus on two representative structures: 
alternating (GC)$_n$ and stacked  G$_n$C$_n$ sequences; the latter structure is argued to
support charge delocalization within G segments, 
and thus an intermediate coherent-incoherent conduction.
We begin our analysis with a highly simplified 1-dimensional tight-binding model, while 
introducing environmental effects through B\"uttiker's probes. This minimal model allows us to
gain fundamental understanding of  transport mechanisms and derive analytic results 
for molecular resistance in different limits.
We then use a more detailed ladder-model Hamiltonian to represent 
double-stranded DNA structures---with environmental effects captured by B\"uttiker's probes.
We find that hopping conduction dominates in alternating sequences, while in stacked sequences 
charge delocalization (visualized directly through the electronic density matrix) supports 
significant resonant-ballistic charge dynamics reflected by an even-odd effect  
and a weak distance dependence for resistance.
Our analysis illustrates that lessons learned from minimal models are helpful for interpreting 
charge dynamics in DNA. 
\end{abstract}

\maketitle

%===============================
\section{Introduction}
\label{Sintro}

% limiting mechanisms
Measurements of charge transfer rates and electrical conductances
in single molecules and self assembled monolayers have revealed 
{\it three} primary limiting transport mechanisms: phase-coherent off-resonance tunneling (superexchange), 
coherent resonant (ballistic) tunneling, 
and incoherent hopping \cite{nitzan,scheer}.
Both on and off-resonance tunneling mechanisms rely on the coherent motion of charges 
through delocalized molecular orbitals.
Deep tunneling conduction decreases exponentially with distance, becoming ineffective in long molecules. 
In contrast, resonant tunneling is insensitive to molecular length, with a thermal
activation profile distinct from that of other thermally-assisted processes such as hopping \cite{Nichols}.
Ballistic motion is often difficult to realize in real molecules, therefore long-range electron transfer 
is typically dominated by incoherent hopping processes, where electrons (or holes) fully localize on molecular sites, 
and hop between them in an incoherent manner.
Such a multi-step hopping conduction is characterized by a linear enhancement of resistance with molecular length.

%DS examples ADD MORE old papers by Ratner
%The two limiting mechanisms, 
Coherent tunneling and multi-step hopping transport mechanisms have been
extensively examined in conjugated molecular wires, see e.g. Refs. \cite{Frisbie1,Frisbie2},
and in biological molecules \cite{Bixon,Barton10,cunibertiR,danny}.
For example, in single DNA molecules, these mechanisms were revealed by studying different
sequences \cite{Kelley,GieseE,GieseR,Tao04,Tao16,Gray}. 
In (GC)$_n$ sequences, site-to-site hopping is the dominant transport mechanism, where
each purine base serves as a hopping site for holes \cite{Tao04,Tao16}. 
In contrast, GC-rich sequences with mediating (A:T)$_m$ blocks support superexchange
with the conductance decreasing exponentially with $m$, and the AT block acting as a tunneling barrier.
Here, A, G, C and T are the adenine, guanine, cytosine and thymine bases, respectively.

%===================================
%Figure 1: scheme
\begin{figure*}[ht]
\vspace{0mm} \hspace{0mm}
{\hbox{\epsfxsize=190mm \epsffile{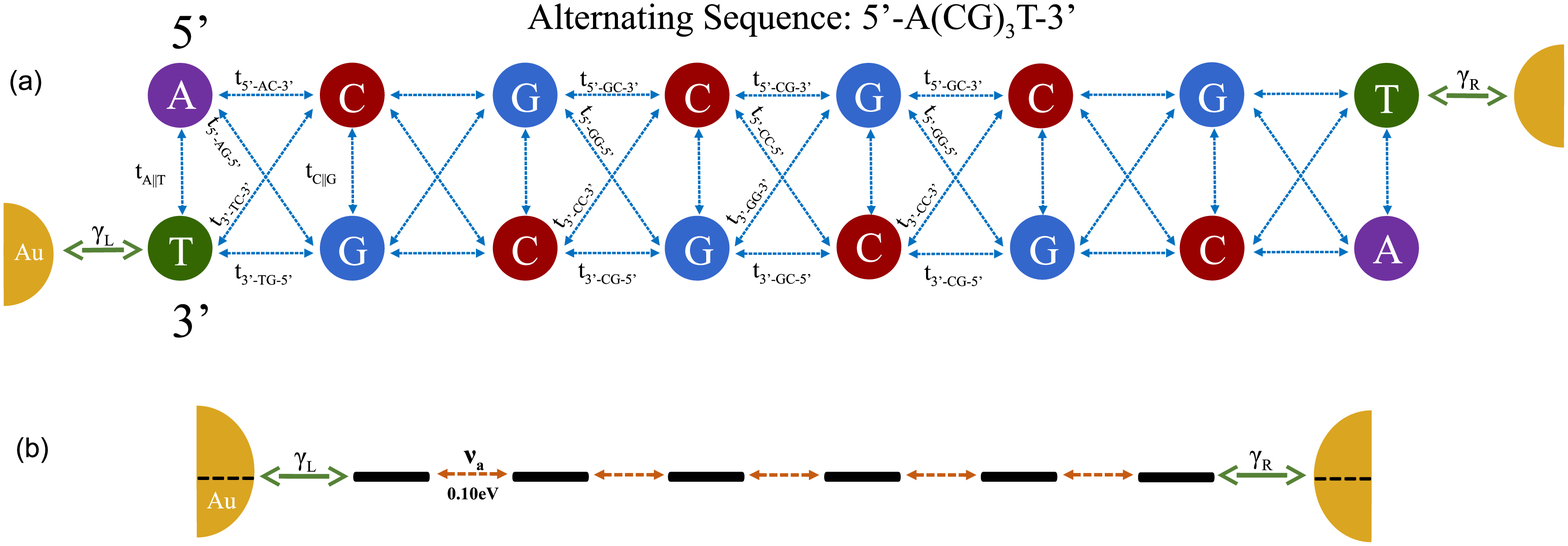}}}
\vspace{10mm}
{\hbox{\epsfxsize=190mm \epsffile{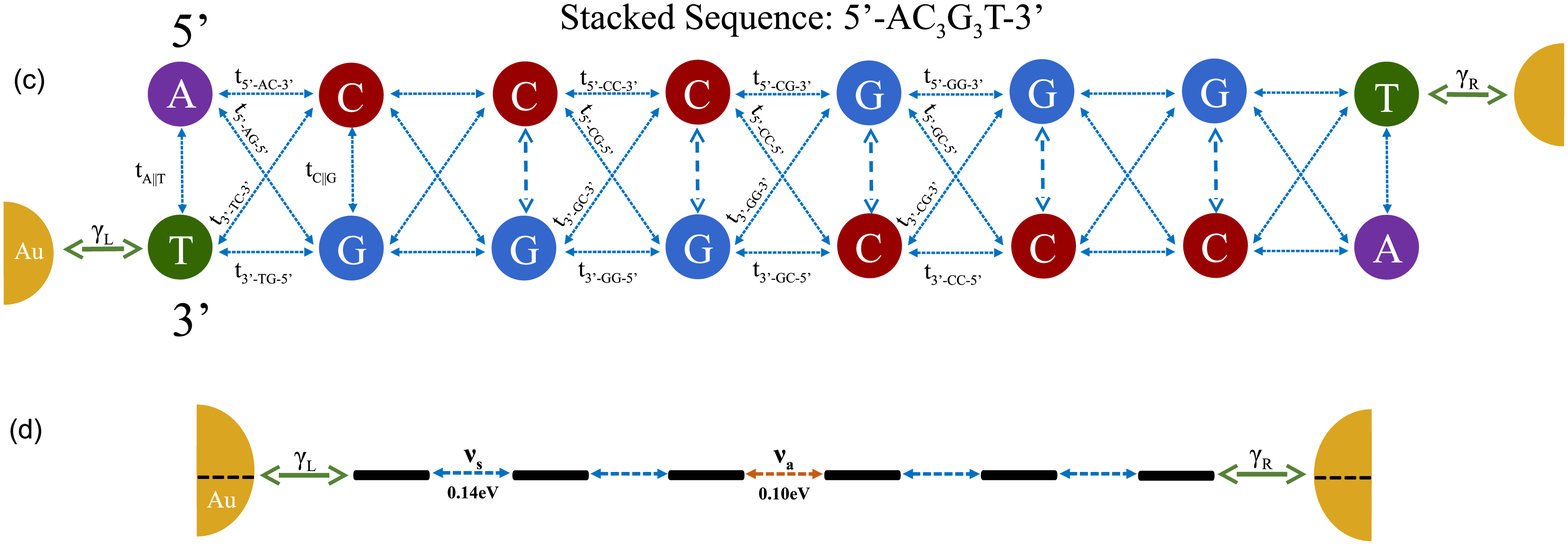}}}
\caption{Example of stacked and alternating DNA sequences investigated in this work.
(a)-(b) and (c)-(d) are the ladder and simplified 1D models for alternating and stacked sequences, respectively. 
See Supporting Information for site energies and tunneling matrix elements. 
}
\label{schemeA}
\end{figure*}
%=========================================

% In between: 
%can it offer more?... long range conduction with lower resistance
Motivated by the challenge to uncover the role of quantum coherent 
effects in biological activity \cite{Scholes}, charge transfer via
intermediate tunneling-hopping conduction has recently received 
theoretical and experimental attention \cite{Berlin13}. 
Particularly, in DNA molecules, an intermediate coherent-incoherent (ICI) situation is expected to be promoted if partial 
delocalization of charges beyond a single G base could be managed.
Indeed, recent measurements of charge transport in double-stranded DNA molecules
suspended between electrodes have revealed an interesting ICI 
transport regime \cite{Ratner15,Tao15}: In an alternating (GC)$_n$ sequence
the resistance increased linearly with length, consistent with the picture of incoherent charge hopping 
between the G sites. In contrast, sequences of two segments of G
bases, G$_n$C$_n$, showed linear enhancement with $n$ dressed by periodic oscillation,
suggesting that charges were partially
delocalized within stacked-G segments \cite{Ratner15}.

% objectives
The objective of the present work is to explore 
intermediate coherent-incoherent charge transport behavior using 
both simple and detailed models relevant for describing conduction in  
DNA junctions.
Our specific objectives are:
(i) Using microscopic models, analyze DNA sequences suggested to promote the intermediate regime and examine
the onset of ICI behavior. % DS
(ii) Study charge delocalization in different sequences directly, 
via the molecular electronic density matrix.
(iii) Suggest simple analytic expressions for describing the intermediate coherent-incoherent regime.
%(iv) Predict structures that are expected to demonstrate significant delocalization behavior.

% Theories
Given the rich electronic structure and complex structural, solvent and counterion dynamics of 
DNA and biomolecules more generally, different computational approaches have been
applied to the study of their charge transport characteristics in different situations.
Hopping transport in DNA was described with kinetic rate equations \cite{Bixon,Burin,danny}, while the
Landauer approach was used to reproduce the tunneling limit using
simple tight-binding models \cite{cuniberti10}. 
Decoherence effects were introduced into charge transport dynamics in DNA using various phenomenological tools \cite{Wolf,probeQi}. % DS
In more sophisticated methods one includes the effect of backbone, solvent, counterions, 
and the DNA internal structural fluctuations by combining classical molecular dynamics (MD) simulations
with quantum mechanics/molecular mechanics methodologies \cite{Beratanrev}.
The resulting coarse-grained electronic Hamiltonian is 
used to compute the transmission function along the MD trajectory \cite{cunibertiJCP09,cunibertiLee,cunibertiNJP10}.
In Ref. \cite{Beratan16}, temporal and spatial fluctuations in the electronic parameters
were introduced based on certain correlation functions. The stochastic Schr\"odinger equation was 
then time-evolved to produce the electronic density matrix and charge current across the system.
In other methods, one explicitly considers the interaction of transport charges with 
selected internal vibrational modes using e.g. Green's function approaches \cite{cunibertiphonon},
quantum rate equations \cite{PeskinPCCP}, or semiclassical approximations \cite{BerlinSC}.

%In this work
In this work, we use a phenomenological tool to implement decoherence and inelastic effects
in molecular junctions, the so called Landauer-B\"uttiker probe (LBP) method \cite{Buttiker-probe1,Buttiker-probe2}. 
In this approach, the non-interacting electronic Hamiltonian is augmented by
probe terminals in which electrons lose their phase-memory
and potentially exchange energy with environmental degrees of freedom (probes). This
technique, originally introduced to study decoherence effects in mesoscopic devices,
has recently been applied to explore electronic conduction in molecular junctions 
\cite{Nozaki1,Nozaki2,Chen-Ratner,WaldeckF,Beratan}.
Particularly, in Ref. \cite{Kilgour1} we demonstrated that the method can capture different
transport regimes in molecular wires: tunneling conduction, ballistic motion, 
and incoherent hopping.
In Refs. \cite{Kilgour2,Kilgour3}, we further used the LBP technique to simulate high-bias voltage effects and
specifically the effect of environmental interactions on diode operation.
In what follows we demonstrate that beyond tunneling and hopping, the LBP framework can further uncover an 
intermediate quantum coherent-incoherent transport regime, with predictions in qualitative 
agreement with recent measurements of charge transport in single-molecule DNA junctions \cite{Ratner15,Beratan16}.

It should be mentioned that transport calculations on double-stranded 
DNA including phenomenologically-introduced decoherence effects
via the LBP framework were previously reported in Ref. \cite{probeQi}.
Our calculations below similarly employ the LBP technique, yet unlike Ref. \cite{probeQi},
we do not rely on the D'Amato-Pastawski effective transmission formula \cite{Pastawski} which is 
limited to the low temperature off-resonance regime. This generalization (or more accurately, 
use of the original LBP equations), allows us to analyze
metal-molecule-metal structures with the Fermi energy positioned in resonance (within $k_BT$) with molecular electronic states.

The paper is organized as follows. In Sec. \ref{Method}, we present the LBP technique.
In Sec. \ref{1D} we introduce a simplified 1-dimensional (1D) model
for charge conduction in alternating or stacked sequences, present simulation results, and 
develop an analytic expression for the ICI regime.
A detailed modeling of alternating and stacked double-stranded DNA molecular junctions 
is included in Sec. \ref{3D} using a ladder model.
We conclude in Sec. \ref{Summ}.

%================================================
\section{Method and System of interest}
\label{Method}

\subsection{B\"uttiker's Probe technique}
\label{method1}

The LBP method allows us to introduce environmental effects into charge transport calculations, including
decoherence, energy exchange, and dissipation. We begin with a general presentation of this method.
Specifically, we use here the so-called voltage probe method  \cite{Buttiker-probe1,Buttiker-probe2} as implemented
in Ref. \cite{Kilgour1}.

The molecular structure is described by a tight-binding Hamiltonian with $j=1,2,..,N$ sites.
The Hamiltonian includes static information on energies of the charge at each site, %nucleobase, and  
and charge transfer integrals between molecular orbitals centered on different sites. %nucleobases.
Below we employ $\nu$ to identify the $L$ and $R$ 
metal electrodes to which the molecule is physically connected.
We count the probe terminals with the index $j$, and use $\alpha$ to identify
all leads: the two metal electrodes $\nu=L,R$  and the $j=1,2,..,N$ probes, acting on each site. 
The total charge current, leaving the $L$ contact, is given by
\bea
I_L=\frac{e}{2\pi\hbar}\sum_{\alpha} \int_{-\infty}^{\infty}
\mathcal T_{L,\alpha}(\epsilon) \left[f_L(\epsilon)-f_{\alpha}(\epsilon)\right]d\epsilon.
\label{eq:currL}
\eea
Here, $f_{\alpha}(\epsilon)=[e^{\beta(\epsilon-\mu_{\alpha})}+1]^{-1}$ is 
the Fermi function in the electrodes, given in terms of the
inverse temperature $k_BT=\beta^{-1}$ and chemical potentials $\mu_{\alpha}$.
The probe functions $f_{j}(\epsilon)$ are determined from the probe condition, to be explained below.
The transmission functions in Eq. (\ref{eq:currL}) are obtained from
the molecular Green's function and the hybridization matrices \cite{nitzan},
\bea
\mathcal T_{\alpha,\alpha'}(\epsilon)={\rm Tr}[ \hat \Gamma_{\alpha}(\epsilon)\hat G^r(\epsilon)\hat \Gamma_{\alpha'}(\epsilon)\hat G^a(\epsilon)].
\label{eq:trans}
\eea
The trace is performed over the $N$ molecular states. The retarded Green's function is given by
\bea
 \hat G^r(\epsilon)=[\hat I\epsilon-\hat H_{M} + i \hat \Gamma/2]^{-1},
\label{eq:Gr}
\eea
with $\hat G^a(\epsilon)=[\hat G^r(\epsilon)]^{\dagger}$, 
$\hat \Gamma=\hat \Gamma_L+\hat \Gamma_R+\sum_{j=1}^{N}\hat \Gamma_j$, 
and $\hat H_M$ the Hamiltonian of the $N$-state molecular system.
In relevant structures, the molecule is coupled to each metal lead through a single site,
with the left (right) lead coupled to site '1' ('$N$').
The $L,R$ hybridization matrices therefore include a single nonzero value,
\bea
[ \hat \Gamma_{L}]_{1,1}= \gamma_L, \,\,\,\,
[\hat \Gamma_{R}]_{N,N}= \gamma_R,
\eea
with $\gamma_{L,R}$  energy parameters describing the metal-molecule coupling.
We work in the wide-band limit: We take
$\gamma_{L,R}$ as energy independent parameters and ignore energy shifts of electronic states due to the leads.
The hybridization matrices $\hat \Gamma_j$ describe the coupling of the $j$th probe to the respective site.
For simplicity, we assume that incoherent effects are local and uniform %acting on electrons localized at each site
with $\gamma_d/\hbar$ as the incoherent rate constant due to environmental processes,
\bea
[ \hat \Gamma_{j}]_{j,j}= \gamma_{d},  \,\,\,\ j=1,2,..,N
%\nonumber\\
\eea
%
%
% Probes
The probes do not only introduce a (trivial) broadening of molecular states, as indicated by 
Eq. (\ref{eq:Gr}). Rather, the voltage probe condition---which we now explain---embodies
incoherent scattering effects, going beyond the coherent Landauer picture.
At finite coupling strength $\gamma_d$, conducting electrons can tunnel
from the molecular system to the probes. % where they lose phase information.
While we enforce the condition of zero leakage current to each probe---electrons lose their phase information in the probes, and they 
re-enter the molecule with a range of energies, determined by the probe energy distribution.

To enforce charge conservation between source ($L$) and drain ($R$), 
the current leaking to each probe is made to nullify. These
$N$ constraints result in $N$
equations for the probes' chemical potentials $\mu_j$. 
In linear response (small bias) these equations can be organized as follows,
\bea
&&\mu_j \sum_{\alpha}\int_{-\infty}^{\infty}\left(-\frac{\partial f_{eq}}{\partial \epsilon}\right)
\mathcal T_{j,\alpha}(\epsilon) d\epsilon
\nonumber\\
&&-
\sum_{j'}\mu_{j'} \int_{-\infty}^{\infty}\left(-\frac{\partial f_{eq}}{\partial \epsilon}\right)
\mathcal T_{j,j'}(\epsilon)d\epsilon
\nonumber\\
&&= \int_{-\infty}^{\infty}d\epsilon
\left(-\frac{\partial f_{eq}}{\partial \epsilon}\right)[\mathcal T_{j,L}(\epsilon)\mu_L
+\mathcal T_{j,R}(\epsilon) \mu_R],
\label{eq:vol1}
\eea
with the Fermi function evaluated at the Fermi energy $\epsilon_F$ (taken as the reference point in our calculations).
We solve these equations for $\mu_j$, the chemical potentials for each probe. In the next step  the set $\mu_j$ is 
used in the calculation of the net current flowing across the device, from $L$ to $R$, 
after linearizing Eq. (\ref{eq:currL}),
\bea
I=\frac{e}{2\pi\hbar}\sum_{\alpha} \left[\int_{-\infty}^{\infty}
\mathcal T_{L,\alpha}(\epsilon) \left(-\frac{\partial f_{eq}}{\partial \epsilon}\right) d\epsilon \right](\mu_L-\mu_\alpha).
\label{eq:currLR}
\eea
When the temperature is low such that the Fermi function can be approximated by a step function at the Fermi energy
(with the derivative a delta function), one can further simplify Eq. 
(\ref{eq:currLR}) and
describe transport in terms of an  effective transmission function 
\cite{Pastawski}.
Since we are interested here in situations in which the temperature is comparable to molecular electronic parameters,
we refrain from making this simplification.

%=================================
\subsection{Molecular system: stacked and alternating sequences}
\label{method2}

Our objective is to explore charge transport behavior in molecular structures
predicted to exhibit intermediate coherent-incoherent conductance characteristics.
Following Ref. \cite{Ratner15}, we focus on two representative DNA sequences:
alternating-G sequences 5'-A(CG)$_n$T-3', denoted by `A', 
and stacked-G sequences, 5'-AC$_n$G$_n$T-3', identified by `S', with $n=2-9$.
The DNA molecules are double-stranded and self-complementary, so as the
complementary sequences of the alternating (stacked) G sequences similarly includes alternating (stacked) G strands.

Since each purine base serves as a hopping site for a hole,
the alternating system is expected to support a localized hopping conduction.
In contrast, conduction in stacked sequences should rely on partial hole
delocalization.
Figure \ref{schemeA} exemplifies the A and S structures for $n=3$. % along with relevant modeling and parameterization.

In Sec. \ref{1D}, we model transport characteristics in the A and S systems 
using a highly simplified 1D model, as we seek to gain basic understanding
of conduction mechanisms in these structures. In Sec. \ref{3D} 
we use a ladder-model Hamiltonian for DNA with input from 
first-principle parametrization \cite{BerlinJacs}.
Both models reveal that, in general agreement with experimental results, resistance in an A-type system increases 
monotonically-linearly with size, indicating on a hopping contribution to the current. %MK
In contrast, S-type molecules conduct through a different mechanism, with greater emphasis on ballistic tunneling.
%while preserving coherences, as 
This is manifested by an even-odd effect of resistance with size, and the resistance's weaker dependence on length.

%--------------------------------------------------------------
\section{1-dimensional model}
\label{1D}

\subsection{Hamiltonian}
We use a simple tight-binding model to represent two families of molecular junctions, alternating and stacked,
as displayed in Fig. \ref{schemeA} panels (b) and (d), respectively.
The modeling is rather minimal and generic yet it could serve to draw some conclusions
on transport in relevant DNA sequences, as well as in other organic and biological molecules.
We describe the alternating junction, loosely corresponding to (CG)$_n$, with $N=2n$ hopping sites,
by two molecular electronic parameters:
the nearest-neighbor electronic coupling $v_a$,
and an onsite energy $\epsilon_B$,
\bea
\hat H_M^A=\sum_{j=1}^{2n} \epsilon_B \hat c_{j}^{\dagger}\hat c_j +
\sum_{j=1}^{2n-1}v_a\left( \hat c_{j}^{\dagger}\hat c_{j+1}  + \hat c_{j+1}^{\dagger}\hat c_{j}  \right).
\label{eq:HA}
\eea
Stacked DNA configurations are made of two segments of G bases. Our 1D model for S systems includes 
three molecular electronic parameters: % see Fig. \ref{fig}.
an onsite energy $\epsilon_B$, the electronic matrix element $v_s$---within the stacked sequence, and 
the electronic coupling of the two segments $v_a$, satisfying $v_s>v_a$. 
The S Hamiltonian reads
\bea
\hat H_M^S&=&\sum_{j=1}^{2n} \epsilon_B \hat c_{j}^{\dagger}\hat c_j +
\sum_{j\neq n} v_s \left(\hat c_{j+1}^{\dagger}\hat c_{j}  + \hat c_{j}^{\dagger}\hat c_{j+1}  \right)
\nonumber\\
&+&
v_{a}\left(
\hat c_{n}^{\dagger}\hat c_{n+1} + \hat c_{n+1}^{\dagger}\hat c_n\right).
\label{eq:HS}
\eea
Eqs. (\ref{eq:HA}) or (\ref{eq:HS}) serve as the molecular Hamiltonian $\hat H_M$ in Eq. (\ref{eq:Gr}).
We use the LBP method as explained in Sec. \ref{Method} and 
study the resistance of the junction  $R\equiv (\mu_L-\mu_R)/I_L$ under low voltage biases.

%MK
We use a uniform chain to capture a single dominant conduction pathway.
Specifically, we assign $\epsilon_B=0$.
We also adopt as representative parameters $v_a=0.1$ eV and $v_s=0.14$ eV, 
following Ref. \cite{Ratner15}.
These electronic coupling energies are presumably too high in the context of DNA, 
yet they still provide a qualitative picture of ICI charge conduction in accord with experiments.
The metal-molecule hybridization is taken in the range 0.05 - 0.5 eV. The applied voltage 
is $\mu_L-\mu_R=0.01$ eV.
We work at room temperature ($22\,^{\circ}{\rm C}$)
and play with environmental effects in the range $\gamma_d=0.01-0.1$ eV.
The position of the Fermi energy relative to molecular states is not given to us in Ref. \cite{Ratner15}.
We thus shift its position and examine examples in the range $|\epsilon_F-\epsilon_B|=0-0.4$ eV.

%==========================
% figure 2 
\begin{figure*}[ht]
\vspace{0mm} \hspace{3mm} 
{\hbox{\epsfxsize=185mm \epsffile{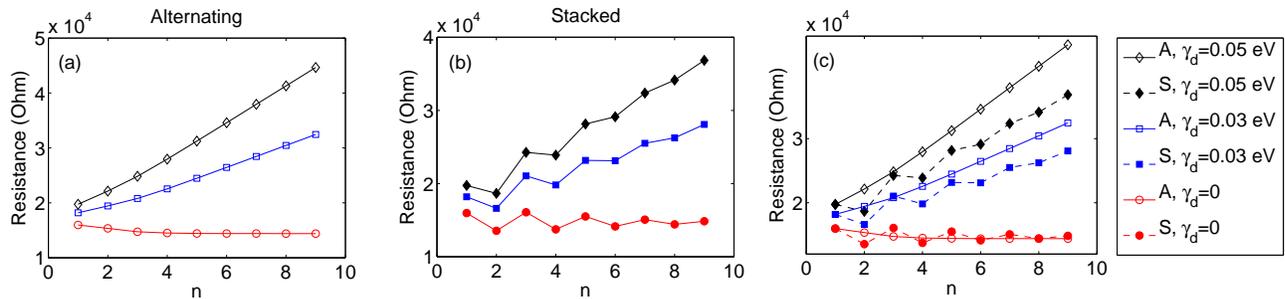}}}   % E0b
\caption{1D Model.
Resistance as a function of length for the (a) alternating  and (b) stacked sequences.
Panel (c) overlays the A and S results.
We use $\epsilon_F=\epsilon_B$, $v_s=0.14$, $v_a=0.1$, $\gamma_{L,R}=0.3$ all in eV, room temperature.
}
\label{FigR1}
\end{figure*}

% figure 3
\begin{figure*}[htbp] 
 \hspace{3mm}
{\hbox{\epsfxsize=185mm \epsffile{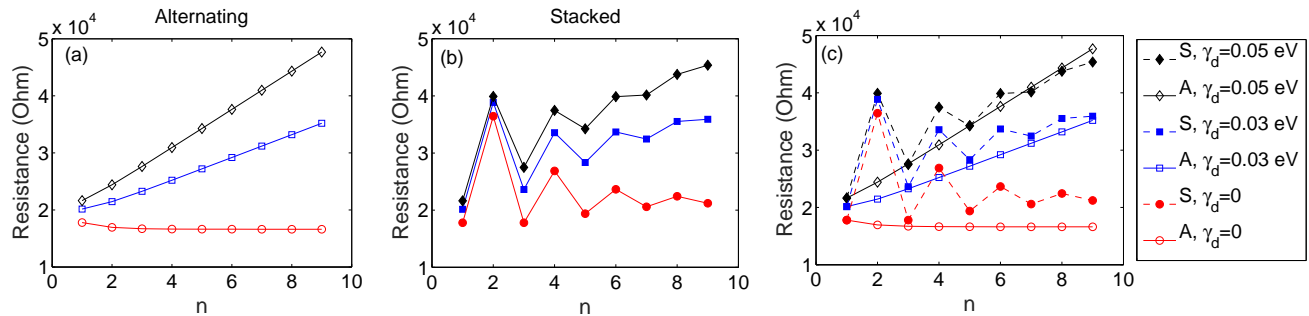}}}  % g1Epn0
\caption{1D Model.
Resistance as a function of length for (a) alternating and (b) stacked  sequences.
Panel (c) overlays the A and S results.
We use the same parameters as in Fig. \ref{FigR1}, with $\gamma_{L,R}=0.1$ eV.
}
\label{FigR2}
\end{figure*}

%============================

\subsection{Simulations: Resistance}

The (GC)$_n$ and the G$_n$C$_n$ sequences constitute $2n$ G sites.
The total number of hopping sites is therefore always even in our simulations, 
but the size of each stacked segment can be even or odd. It is important to mention here that the 1D system we 
present is not a direct analog for DNA, and as such we do not use parameters directly drawn from real DNA
molecules, and thus do not expect to quantitatively retrieve experimental results. 
The system under study is more generally a near-resonance conductor which provides mechanistic
insight into ICI conduction thought to exist in stacked/alternating DNA.

Figure \ref{FigR1} displays the resistance of the A and S molecules in a resonance situation, $\epsilon_F=\epsilon_B$,
using $\gamma_{L,R}=0.3$ eV.
When $\gamma_d=0$,
the two sequences conduct charges via a coherent-resonant (ballistic) mechanism \cite{Kilgour1}.
The alternating sequence shows a distance-independent behavior at large $n$ once residual 
off-resonance contributions become small.
The S sequence displays an oscillatory behavior. Note that
S sequences with an odd number of bases in each segment ($n=1,3,5,..$) 
show higher resistances than segments with even number of stacked bases, $n=2,4,6,..$.
It should be emphasized that A molecules demonstrate parallel even-odd effects---not observed here
since A sequences here are always of an even size. 

%just curious - tells us how "incoherent" the total current is 

Introducing incoherent effects, $\gamma_d\neq0$, Fig. \ref{FigR1} shows that in A molecules 
the resistance grows linearly with size. In contrast,
in the S sequences, an oscillatory behavior is superimposed on the linear enhancement of $R$ with size, surviving up to 
$n\sim 8$. 

In Fig. \ref{FigR2} we repeat this calculation---only with a weaker metal-molecule hybridization
$\gamma_{L,R}=0.1$ eV, and reveal a rather interesting effect: 
S sequences are more significantly affected by $\gamma_{L,R}$ 
as opposed to A-type molecules. Further, the resistance of S junctions increases
upon reducing $\gamma_{L,R}$,
as compared to the respective A sequences.
We can rationalize these observations as follows.
At finite $\gamma_d$, charges in the A sequence primarily %MK
hop site-to-site, thus the overall resistance is rather insensitive 
to the contact energy $\gamma_{L,R}$.
By contrast, in S sequences charges are more delocalized, %MK
thus charge conduction is more susceptible to 
$\gamma_{L,R}$, resulting in a higher overall resistance at smaller hybridization.
A further curious observation is that the even-odd effect has switched parity, with even junctions displaying higher
resistance than odd, unlike in Fig. \ref{FigR1}. We elucidate this effect in the next section.

In the Supporting Information file we include results working under different conditions, when the 
molecular states are shifted away from the Fermi energy. We find that the conductances of the
S and A structures fundamentally differ in other parameter regimes, 
with S structures showing intriguing non-monotonic behavior.

%=====================================

%\subsection{Analytic Results}
%\label{SAnal}

Several interesting questions come to mind when inspecting Figs. \ref{FigR1}-\ref{FigR2}. 
(i) Which sequence should act as a better conductor, S or A?
(ii) In the stacked configuration, when should odd-$n$ segments or even-$n$ segments better conduct?
(iii) The oscillations in the conductance are damped out with $n$, even when $\gamma_d=0$.
 What factors determine this damping, at $\gamma_d=0$, and at $\gamma_d\neq0$? % voltage, temperature
(iv) What mechanisms dominate charge transport in the different sequences?
Can we organize analytic expressions for the resistance per site for the S and A structures?
%How does it depend on environemtnal effects, $\gamma_d$, temperature?
We address these questions by developing limiting analytical expressions for the 
electronic conductance in our model.

%=============================
\subsection{Analytic results}

In this section we develop approximate expressions for the resistance in the S and A structures in 
(i) the fully coherent case, and (ii) in the incoherent regime. 
We then suggest an interpolating expression
which describes intermediate coherent-incoherent results.

We begin by ignoring environmental effects altogether, $\gamma_d=0$. 
Furthermore, for simplicity, we assume here that the temperature is rather low,
$k_BT<v_a,v_s,\gamma_{L,R}$ (our simulations do take into account finite-temperature effects).
We also assume that the junction is coupled identically at the two ends and denote $\gamma=\gamma_{L,R}$.
The transmission function of a resonant, uniform, and symmetrically-coupled bridge of length $N=2n$, 
can be obtained analytically rather easily. 
The alternating sequence supports the transmission probability
\bea
\mathcal T_A^{(\gamma_d=0)}(\epsilon_F)=\frac{\gamma^2v_a^2}{(v_a^2+\gamma^2/4)^2}.
\label{eq:TA}
\eea
%
%where the superscript '0' identifies the coherent limit, $\gamma_d=0$.
Stacked sequences in contrast distinguish between even and odd-length segments,
\bea
\mathcal T_{S,even}^{(\gamma_d=0)}(\epsilon_F)&=&\frac{\gamma^2v_s^4v_a^2}{(v_s^4+v_a^2\gamma^2/4)^2}, \,\,\,\,  n \,\, {\rm even} %n=2j, 
\nonumber\\
\mathcal T_{S,odd}^{(\gamma_d=0)}(\epsilon_F)&=&\frac{\gamma^2v_a^2}{(v_a^2+\gamma^2/4)^2}, \,\,\, n \,\, {\rm odd}%n=2j+1.
\label{eq:TS}
\eea
We can immediately confirm that the inequality % $\mathcal T_{S,even}^{(0)}(\epsilon_F) >\mathcal T_{S,odd}^{(0)}(\epsilon_F) =\mathcal T_{A}^{(0)}(\epsilon_F)$,
\bea
\mathcal T_{S,even}^{(\gamma_d=0)}(\epsilon_F) =
\frac{\gamma^2v_s^4v_a^2}{(v_s^4+v_a^2\gamma^2/4)^2} &>&
\frac{\gamma^2v_a^2}{(v_a^2+\gamma^2/4)^2}
\nonumber\\
&=&\mathcal T_{S,odd}^{(\gamma_d=0)}(\epsilon_F) 
%\nonumber\\
 =\mathcal T_{A}^{(\gamma_d=0)}(\epsilon_F)
\nonumber
\eea
is equivalent to
\bea
(\gamma^2-4v_s^2)(v_s^2-v_a^2)>0
\eea
or (since $v_a<v_s$),
\bea
\gamma>2v_s.
\label{eq:ineq}
\eea
We recall that at low temperatures conductance (inverse of resistance) is 
proportional to the transmission function.
This last inequality thus resolves the first two questions raised above: %at the begining of this section:
(i) Even-length S sequences conduct more effectively (smaller resistance) than A sequences, as long as  $\gamma>2v_s$.
%in line with the results of Fig. \ref{FigR1}.
(ii) The even-odd effect observed for the S sequence is determined by the relative magnitude of $\gamma$ and $v_s$.
When $\gamma>2v_s$, the resistance of odd-length segments (e.g. G$_5$C$_5$) is greater than the resistance of 
even-length segments (e.g. G$_6$C$_6$), in agreement with simulations displayed in  Figs. \ref{FigR1}-\ref{FigR2}.

Eqs. (\ref{eq:TA})-(\ref{eq:TS}) were developed in the resonant-coherent limit. They do not predict damping of the even-odd effect or saturation of resistance for the S sequence at long $n$.
These oscillations are obviously damped out even for $\gamma_d=0$, see Fig. \ref{FigR1}, 
as a result of finite voltage and temperature effects.
These equations further reveal an intriguing phenomenon:
The different sequences conduct identically, 
$\mathcal T_{S}^{(\gamma_d=0)}(\epsilon_F) =\mathcal T_{A}^{(\gamma_d=0)}(\epsilon_F)$, at two points. 
The first solution is trivial,
$v_s=v_a$, when the two sequences are identical.
More peculiar is the solution at $\gamma=2v_s$.
In this case the A and S sequences behave identically in the coherent limit, though their electronic structure
significantly differs.
%aconduct and even-odd oscillations in S sequances disappear.
This unusual behavior highlights the important role of 
the boundary hybridization $\gamma_{L,R}$  on ballistic conduction,
unlike the ohmic-hopping limit discussed below.

\begin{figure}[htbp]
\vspace{0mm} \hspace{3mm}
{\hbox{\epsfxsize=70mm \epsffile{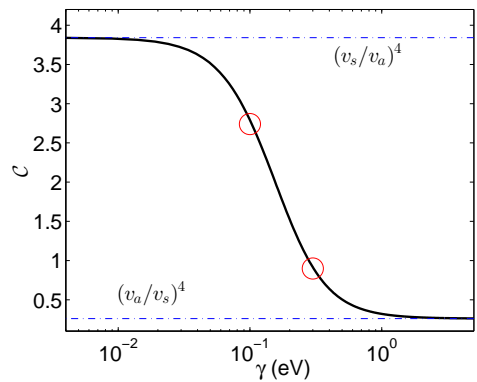}}} % ratio.eps
\caption{1D Model.
Ratio of resistances in even and odd S sequences as a function of metal-molecule coupling energy
$\gamma=\gamma_{L,R}$.
We use $\epsilon_B-\epsilon_F=0$, $v_s=0.14$, $v_a=0.1$  in eV.
Dash-dotted lines identify asymptotic values. 
The two circles mark the values at $\gamma$ used in  Figs. \ref{FigR1} and \ref{FigR2}.
}
\label{FigC}
\end{figure}

What factors determine the contrast of resistances, $\mathcal C=R^{(\gamma_d=0)}_{S,even}/R_{S,odd}^{(\gamma_d=0)}$,
between even and odd-length S sequences? Obviously, we should set $v_s\neq v_a$.
In Fig. \ref{FigC} we further demonstrate that $\mathcal C$
can be tuned by the metal-molecule hybridization, reaching 
the asymptotic values $(v_a/v_s)^{\pm 4}$ at low and high hybridization.
Within our parameters, $(v_s/v_a)^2\sim 4$. Therefore, by manipulating contact energy 
we could in principle 
reach high contrasts up to a factor of 4. 
From Fig. \ref{FigC} we learn that
the value $\gamma=0.3$ eV, used in our simulations, lead to a mild contrast 
$\mathcal C\sim 1.2$, in agreement with trends observed in Fig. 
\ref{FigR1}(b) (recall, $\gamma_d=0$).

Eq.  (\ref{eq:TS}) indicates that by weakening the coupling energy 
between stacked  segments $v_a$, 
we should obtain enhanced resistance oscillations as a function of length $n$. 
This observation agrees
with measurements and calculations reported in Ref. \cite{Beratan16} 
where  the coupling between the stacked blocks was controlled by swapping the 3'5' orientation 
of the sequence with respect to the metal leads.

%==========================
% figure 4 Ratio

% Figure 5 fit
\begin{figure}[htbp]
%\vspace{-32mm}
 \hspace{3mm}
{\hbox{\epsfxsize=90mm \epsffile{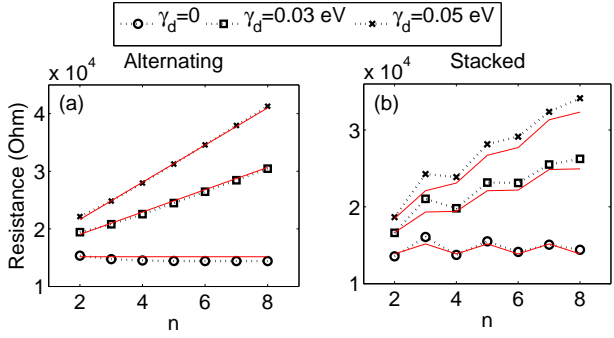}}} % fittB.eps
\caption{
1D Model.
Fitting numerical simulations (symbols) by approximate analytical expressions (full lines), 
Eqs. (\ref{eq:RAf}) and (\ref{eq:RSf})
for the A and S sequences, respectively.
Parameters correspond to Fig. \ref{FigR1}.
}
\label{Figfit}
\end{figure}
%============================

We now turn our attention to the fully incoherent-ohmic limit. The 
following expression can well describe the slope in Figs. \ref{FigR1}-\ref{FigR2},
\bea
R_{ohm}\propto\frac{\gamma_d}{4|v|},
\label{eq:Rohm}
\eea
with $v=v_{a,s}$ for the A and S sequences, respectively.
We can rationalize this expression as follows. For long enough molecules, $n>5$,
the eigenenergies of the tight-binding Hamiltonian form a band of width
$4|v|$. % $v=v_{a,s}$ for the A and S sequences, respectively.
This band picture is valid for the 
S configuration as well. % since the lattice is almost periodic as an A molecule.
At every site, electrons suffer incoherent effects of strength $\gamma_d$.
The resistance per site is given by the competition between the energy scale for inelastic effects
which determine site-to-site hopping, $\gamma_d$, and the bandwidth $4v$.

Using the two limiting expressions for the fully coherent and the incoherent scenarios,
Eqs. (\ref{eq:TS})-(\ref{eq:TA}) and (\ref{eq:Rohm}), respectively,
we  organize an interpolating expression for the total resistance, 
to describe ICI situations. For the A sequence we write
\bea
R_A\sim\frac{1}{G_0}\frac{\gamma_d}{4|v_a|} N +  \frac{1}{G_0}\frac{(v_a^2+\gamma^2/4)^2}{\gamma^2v_a^2}, 
\label{eq:RAf}
\eea
with $N=2n$ as the number of sites, $G_0$ is the quantum of conductance. 
For the S sequence we suggest the form
\bea
R_S\sim\frac{1}{G_0}\frac{\gamma_d}{4|v_s|} N +%+  \frac{1}{G_0}\frac{(v_a^2+\Gamma^2/4)^2}{\Gamma^2v_a^2}
\begin{cases}
\frac{1}{G_0}\frac{(v_a^2+\gamma^2/4)^2}{\gamma^2v_a^2}, &\text{ odd } n \\
\frac{1}{G_0}\frac{(v_s^4+v_a^2\gamma^2/4)^2}{\gamma^2v_s^4v_a^2}, & \text{ even } n.\\
\end{cases}
\nonumber\\
\label{eq:RSf}
\eea
The different terms in Eqs. (\ref{eq:RAf}) and (\ref{eq:RSf}) describe the competition between transport mechanisms:
hopping conduction (first term) and ballistic transport---which is determined by the contact energy (second term). 
Since hopping resistance grows linearly with molecular length, it dominates charge transfer in long structures.

In Fig. \ref{Figfit} we demonstrate that the interpolating expressions reasonably fit our 1D 
simulations. 
We note that in the S sequence the resistance per site is slightly greater than the fitted slope of $\gamma_d/|4v_s|$,
indicating that we should probably use an effective-smaller bandwidth in analytical calculations of S molecules.
This makes sense given the weak link at the center between segments.
Indeed, simulation results were better reproduced by analytic expressions
with $\sim 0.9 v_s$ as a measure the for the electronic coupling in the tight-binding lattice.
Note that our fitting expressions do not include the effect of temperature and bias voltage, 
which are responsible for  damping out coherent oscillations. 

Our conclusion from this fitting analysis is that both structures, A and S, 
preserve the effect of ballistic-coherent conduction under environmental effects---when
$n=2-8$. 
However, while in the case of A sequences the coherent contribution to the total resistance 
is rather trivial, a constant factor,
in S-type structures the even-odd nature of the coherent term manifests itself as a nontrivial oscillatory
contribution---on top of the ohmic resistance.

%=====================================
% figure 6 1D RDM 
\begin{figure*}[htp]
\vspace{-0mm} \hspace{-10mm}
{\hbox{\epsfxsize=86mm \epsffile{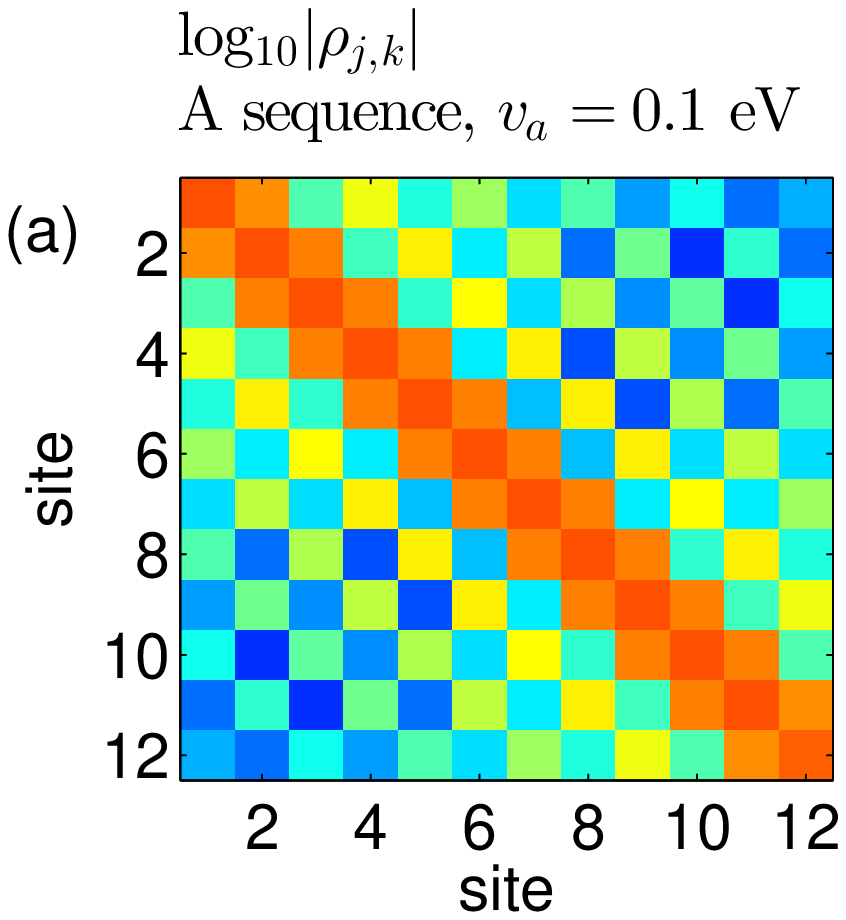} % RDM1aa
\hspace{-32mm}
 \epsfxsize=86mm \epsffile{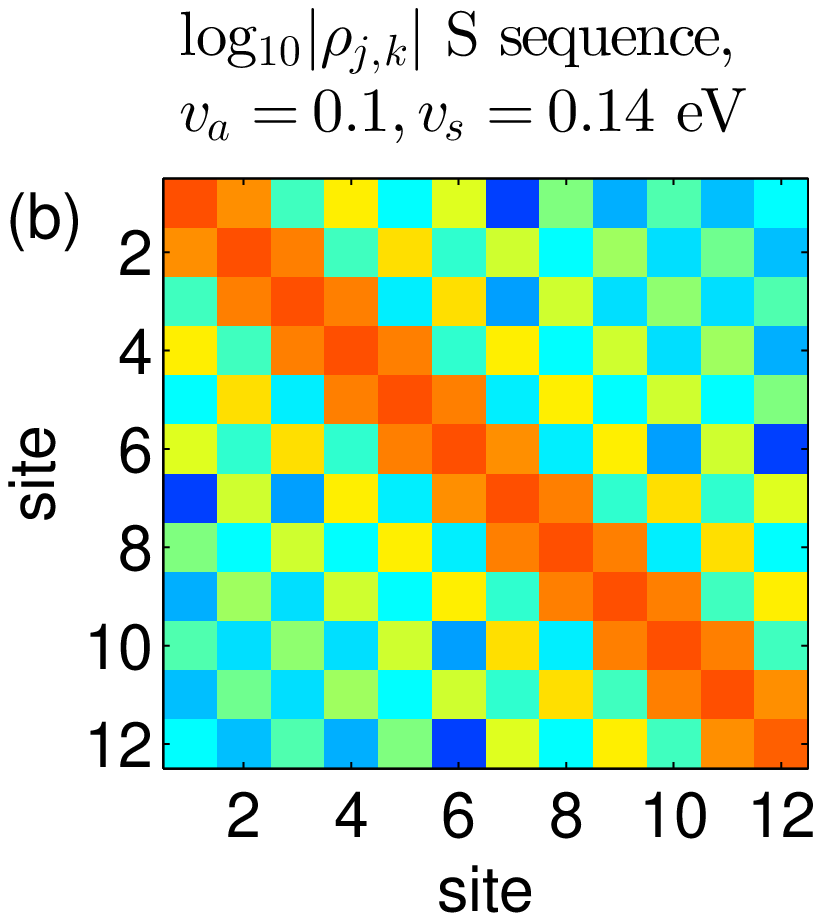} % RDM1bb
\hspace{-32mm} 
\epsfxsize=86mm \epsffile{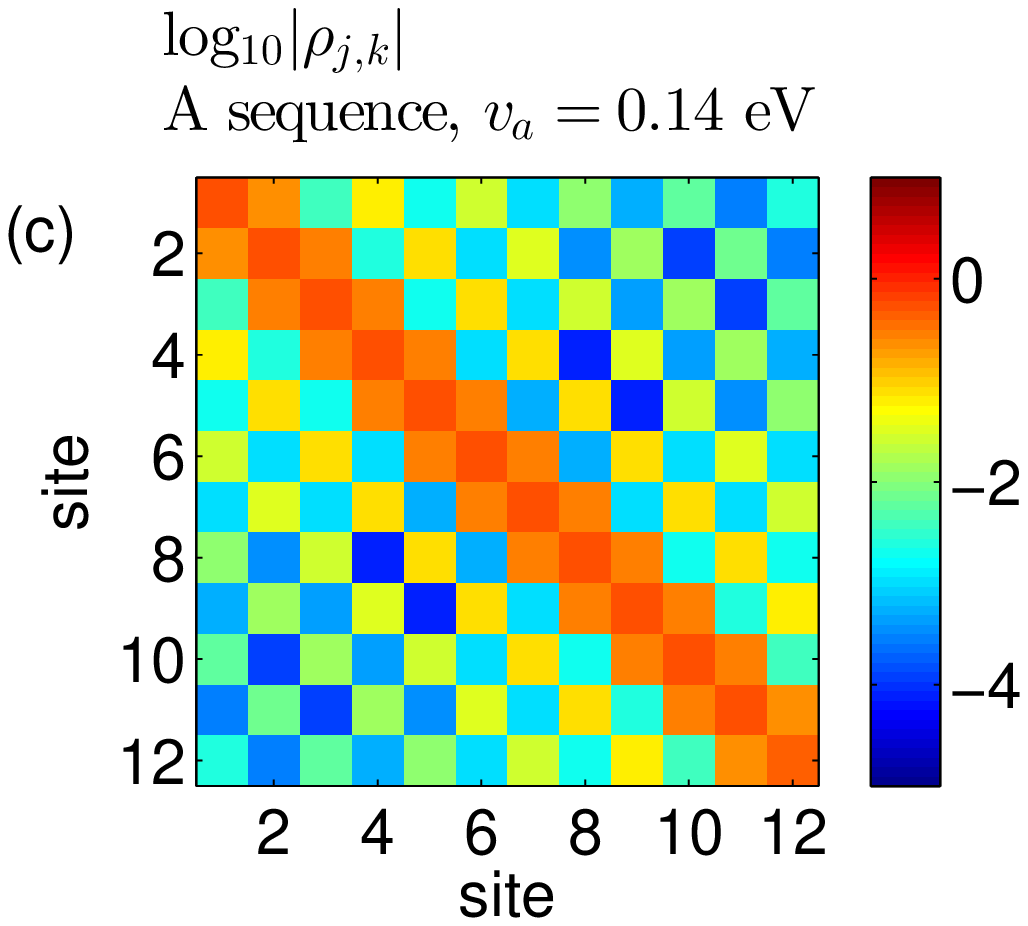}} % RDM1cc
}
\caption{1D Model:
Coherence properties of alternating and stacked sequences
manifested through a 
density matrix map (absolute values) for $n=6$ chains.
(a) A sequence, %with $v_a=0.1$ eV, 
(b) S sequence %with $v_a=0.1$ and $v_s=0.14$ eV,
and (c) an A sequence with an enhanced electronic matrix element. %, $v_a=0.14$ eV.
We use $\epsilon_F=\epsilon_B$, $\gamma_d$=0.05, $\gamma_{L,R}=0.3$ in eV, room temperature.
}
\label{Figrdm1}
\end{figure*}
%=====================================

%figure 7
\begin{figure*}[htp]
\vspace{-4mm} %\hspace{-26mm} 
{\hbox{\epsfxsize=86mm \epsffile{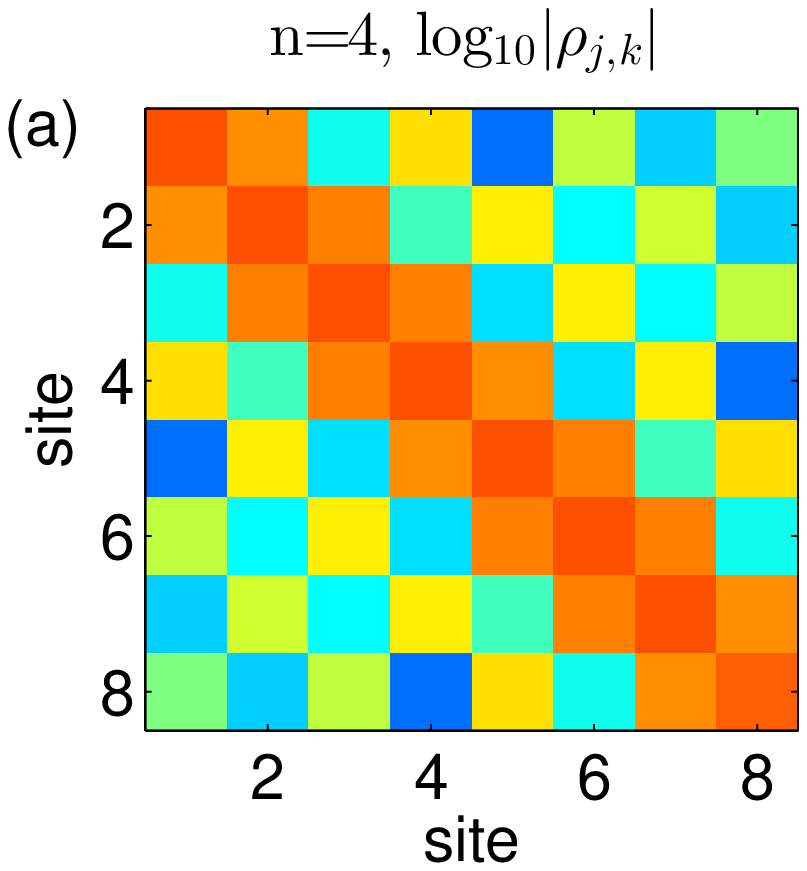} % RDM2aa.eps
\hspace{-34mm}
\epsfxsize=86mm \epsffile{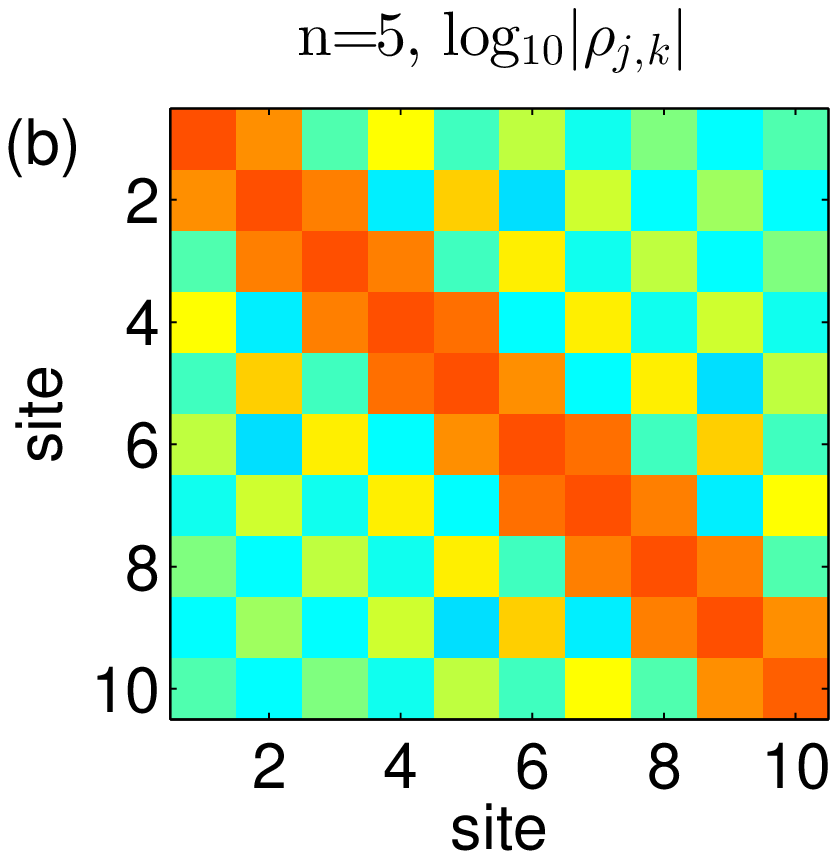} % RDM2bb.eps
\hspace{-34mm}
\epsfxsize=86mm \epsffile{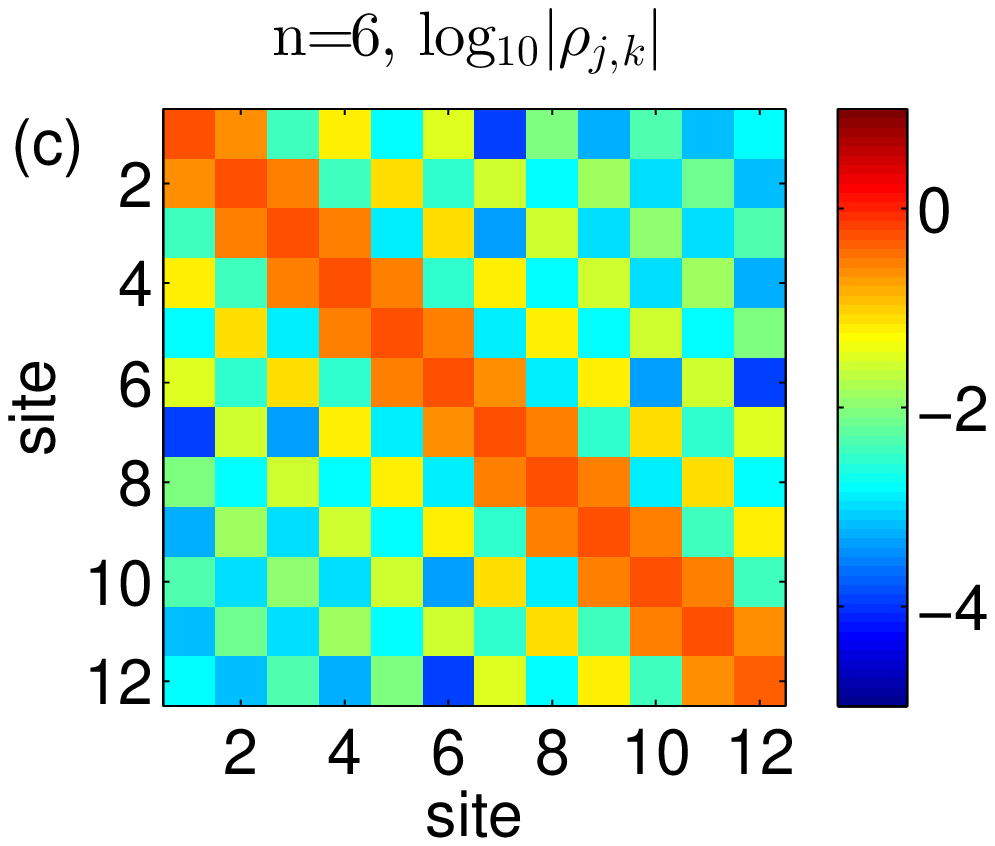}} % RDM2cc.eps
}
\caption{1D Model:
Coherence properties of even and odd Stacked sequences manifested through a
density matrix map (absolute values) of stacked chains of (a) $n=4$, (b) $n=5$, and (c) $n$=6.
$\epsilon_B=\epsilon_F=0$, $\gamma_d=0.05$, $\gamma_{L,R}=0.3$, all in eV,  room temperature.
}
\label{Figrdm2}
\end{figure*}

%%=================================

\subsection{Density matrix}

The steady state electronic density matrix of the molecule contains information on the 
stationary population (diagonal elements)
and surviving coherences between sites (off-diagonal terms).
The density matrix can be readily calculated within the LBP method \cite{salil}
%\cite{Dhar}, 
%
\bea
% <c_k^{\dagger}c_j> check
\rho_{j,k}=\frac{1}{2\pi}\sum_{\alpha}\int_{-\infty}^{\infty} d\epsilon [\hat G^r(\epsilon)\hat\Gamma_{\alpha}\hat G^a(\epsilon)]_{j,k}f_{\alpha}(\epsilon),
\eea
 with the probe distribution functions as received from the probe condition.

In Fig. \ref{Figrdm1} we investigate the coherence properties of A and S chains for a certain size, $n=6$.
The population is about constant (close to 0.5) all through,
coherences display an even-odd effect.  
We note that coherences are significantly greater in the S system (panel b) as compared to the A molecule (panel a),
indicated by the light blue color on the off-diagonals in the former, compared to the dark blue color in panel (a).
Why do coherences better survive in the S configuration? 
Is it because of the imbalance in 
electronic coupling, $v_s>v_a$, or is the effect related to the two-segment 
configuration of S molecules?
Panel (c) points to the latter explanation. We
study there an alternating chain while increasing the electronic coupling to $v_a=0.14$ eV (equal to $v_s$).
Comparing results from panels (b) and (c), we find that the stacked configuration (b) supports coherences
that are greater than those in a tightly-coupled yet uniform chain. 

It was argued in Ref. \cite{Beratan16} that in an odd-$n$ G block, coherent charge transport is established and
in contrast, even-length G systems support incoherent transport.
In Fig. \ref{Figrdm2} we demonstrate that odd-$n$ stacked structures indeed maintain strong coherences---larger
than even-$n$  stacked structures. 
However, both even and odd-length $n$ blocks 
show pronounced coherent properties---stronger than alternating structures---as was demonstrated
in Fig. \ref{Figrdm1}. %MK - for these params, the even S conducts better anyway, despite lower coherences
Thus, we argue that stacked sequences, of both even and odd length, conduct 
via a partially coherent mechanism.

%=====================================

\section{Ladder model for double-stranded DNA}
\label{3D}

In the previous section we had demonstrated that A and S 
chains support different transport characteristics:
In A-type structures the resistance grows monotonically-linearly,  while
S sequences evince an even-odd coherent effect that is visible 
on top of the linear-ohmic resistance. 
These observations were drawn from a simple 1D tight-binding Hamiltonian.
Is this simple picture appropriate for describing double-stranded (ds) DNA structures, notorious for their complexity?
In this section, we model A and S sequences %of Ref. \cite{Ratner15}
by a tight-binding ladder-model Hamiltonian, see e.g. Ref. \cite{cunibertiphonon, ladder1,ladder2,ladder3,Wolf}. 
The ladder Hamiltonian describes the topology of a ds-DNA molecule which is $n$ base-pairs long,
with each site representing a particular base.
We assume that charge transport takes place along the base-pair stacking, ignoring the backbone,
\bea
\hat H_M&=&\sum_{j=1}^n \Bigg[
\sum_{s=1,2} \epsilon_{j,s}\hat c_{j,s}^{\dagger}\hat c_{j,s} % energies
+ \sum_{s\neq s'=1,2} t_{j,ss'}\hat c_{j,s}^{\dagger}\hat c_{j,s'}
\nonumber\\
&+&
 \sum_{s,s'=1,2} t_{j,j+1,ss'}(\hat c_{j,s}^{\dagger}\hat c_{j+1,s'} + h.c.) \Bigg].
\eea
% n=(N-4)/4
%
The index $s=1,2$ identifies the strand. $\hat c_{j,s}^{\dagger}$ creates a hole on strand $s$ at the $j$th site with the
on-site energy $\epsilon_{j,s}$, $t_{j,ss'}$ and $t_{j,j+1,ss'}$
are the electronic matrix elements between nearest neighboring bases.
This model mimics the topology of the ds-DNA molecule; helical effects are effectively 
included in the renormalized electronic parameters.

We use the parametrization of Ref. \cite{BerlinJacs}, which is calculated at the  DFT level,
distinguishing between backbone orientations (5' and 3').
In this parametrization, on-site energies vary depending on the identity of neighboring sites.
Here, following Ref. \cite{Wolf}, we simplify this description and assign a single value (averaged) 
for on-site energies for each base,
see Table I. The electronic matrix elements were taken directly from Ref. \cite{BerlinJacs},
for completeness, tables are included in the Supporting Information. % DS

%================================================================================================
% Table I
\vspace{4mm}
\label{tab:title}
\begin{center}
Table I: Selected on-site energies and inter-strand electronic coupling (eV) \cite{BerlinJacs,Wolf}. 
(See the supporting information for full parameter set.)\\ %MK hold for more specific reference to supporting
\begin{tabularx}{.46\textwidth} { c c  c  c c c  } 
\hline
\hline
$\epsilon_G$ \hspace{3mm} & $\epsilon_A$ \hspace{3mm} & $\epsilon_C$ \hspace{3mm} & $\epsilon_T$ \hspace{3mm} & $t_{\rm{G||C}}$ \hspace{3mm} &  $t_{\rm{A||T}}$\\
\hline
8.178 \hspace{3mm} & 8.631 \hspace{3mm} & 9.722 \hspace{3mm}& 9.464 \hspace{3mm} & -0.055 \hspace{3mm}& -0.047 \\
\hline
\hline
\end{tabularx}\par
\end{center}
\vspace{4mm}

%%%%%%%%%%%%%%%%%%%%%%%%%%%%%%%%%%%%%%%%%%%%%%%%%%%

%=====================================
%figure 8 ladder1
\begin{figure*}% [htbp]
\vspace{0mm} \hspace{0mm}
{\hbox{\epsfxsize=170mm \epsffile{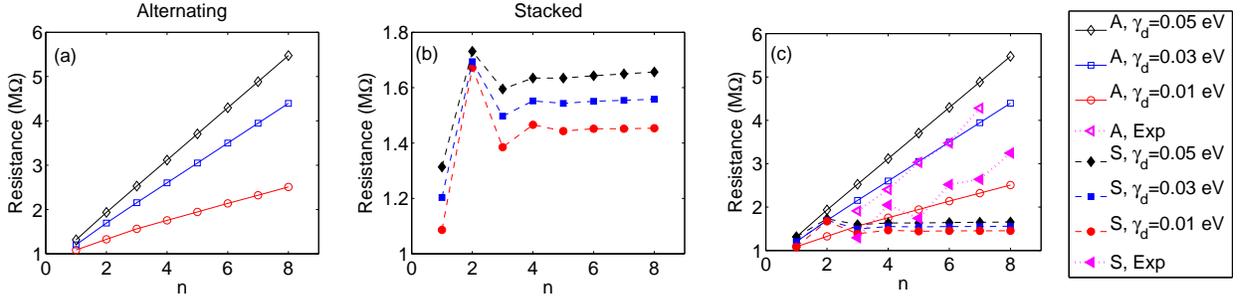}}}  % ladder1N.eps
\caption{Ladder model.
Resistance of (a) alternating (GC)$_n$ and (b) stacked G$_n$C$_n$ sequences,
see Fig. \ref{schemeA} for the ladder model.
Panel (c) overlays A and S simulations along with experimental results from Ref. \cite{Ratner15}.
Simulations were performed at room temperature, with $\gamma_d=0.01$, 0.03 and 0.05 eV, 
$\gamma_{L,R}=0.05$ eV.
}
\label{ladder1}
\end{figure*}

% figure 9 ladder2
\begin{figure*}[htbp]
\vspace{0mm} \hspace{0mm}
{\hbox{\epsfxsize=170mm \epsffile{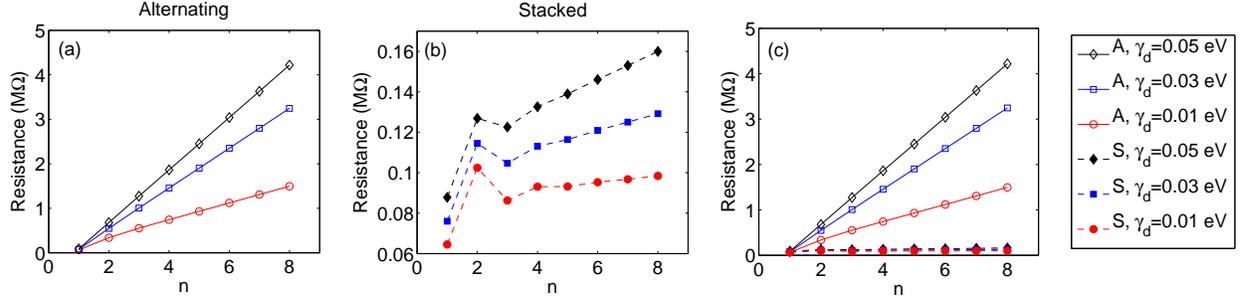}}} % ladder2N
\caption{Ladder model.
Resistance of (a) alternating (GC)$_n$ and (b) stacked G$_n$C$_n$ sequences,
panel (c) overlays A and S results.
Parameters are the same as in Fig. \ref{ladder1}, but we take $\gamma_{L,R}=1$ eV.
}
\label{ladder2}
\end{figure*}

%=================================

% figure 10
\begin{figure}[htbp]
\vspace{0mm} \hspace{-10mm}
{\hbox{\epsfxsize=90mm \epsffile{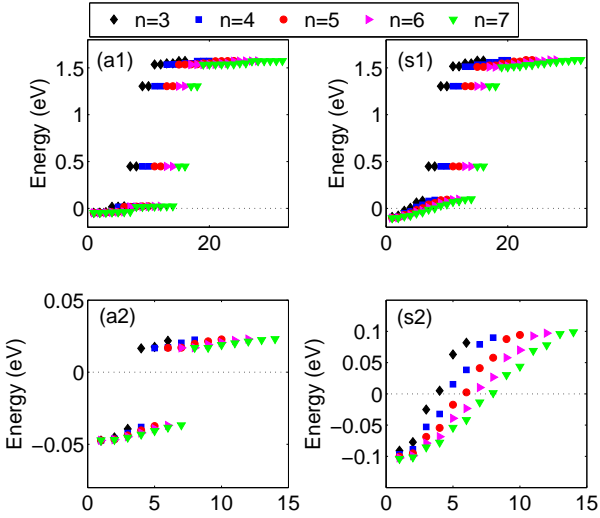}}}  % energy
\caption{Energy spectrum 
for alternating (a1)-(a2) and stacked (s1)-(s2) ladder models,
with $N=4n+4$ electronic states.
Panels (a2) and (s2) zoom over the low-energy range.
The dashed line marks the Fermi energy, and we set 
$\epsilon_F=\epsilon_G=0$.
}
\label{eigg}
\end{figure}

%=====================================

% figure 11
\begin{figure*}[htp]
\vspace{-0mm} \hspace{-10mm}
{\hbox{\epsfxsize=93mm \epsffile{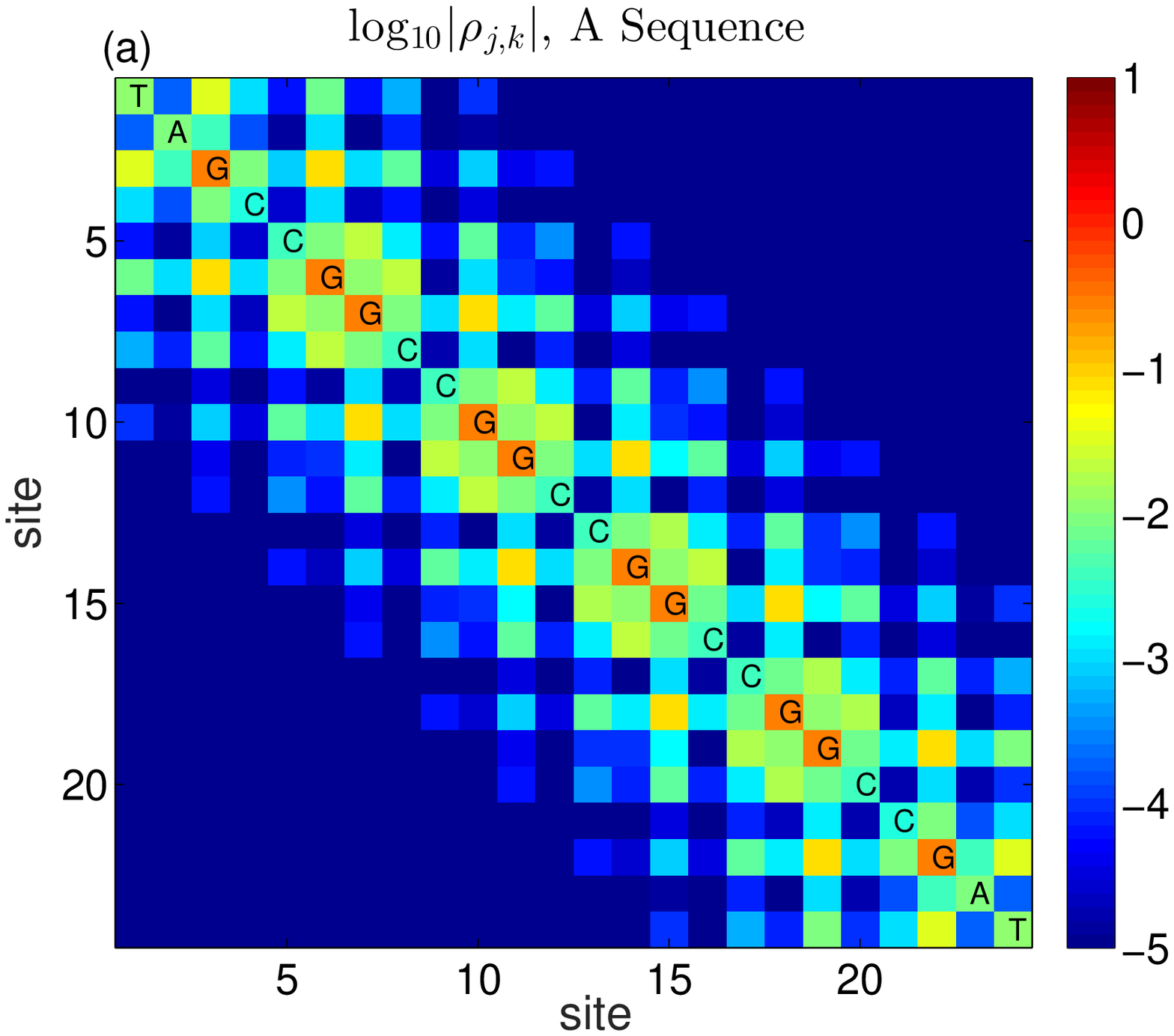}   % A24.eps
\hspace{-10mm}
\epsfxsize=93mm \epsffile{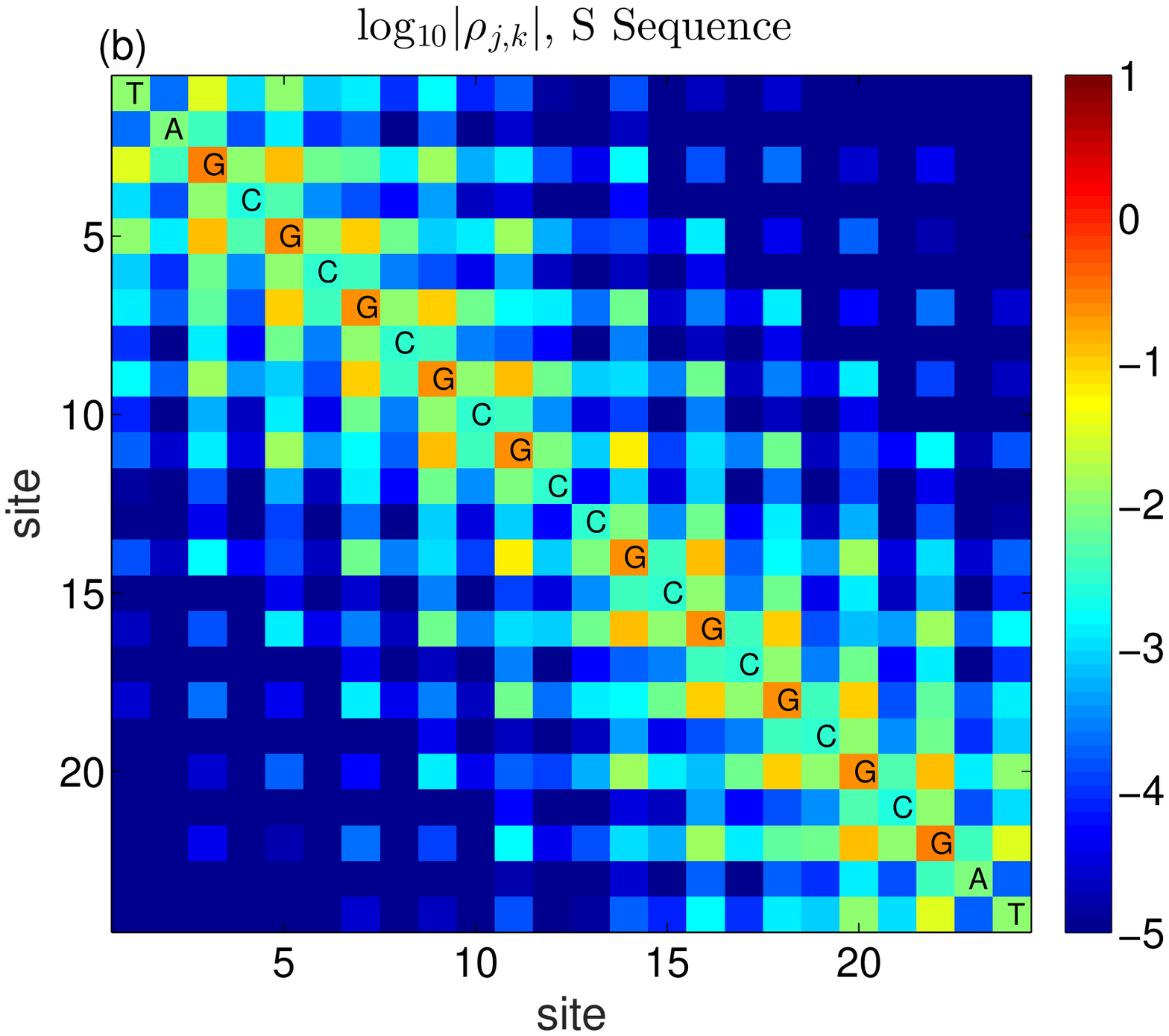}}   % S24.eps
}
\caption{Density matrix map of the ladder model for
 (a) alternating  and (b) stacked sequences.
Parameters are
$\gamma_{L,R}=0.05$ eV, $\gamma_d=0.03$ eV, room temperature,
$\epsilon_F=\epsilon_G$.
}
\label{DNArdm1}
\end{figure*}
%=====================================

Given the molecular Hamiltonian, we calculate transport characteristics of the metal-molecule-metal
junction using the LBP method as described in Sec. \ref{Method}.
We connect the DNA molecule to metal leads as sketched in Fig. \ref{schemeA}, and take into account
environmental effects (structural motion, solvent, counterions)
by attaching local probes to each site.
Following experiments \cite{Ratner15,Beratan16}, our simulations explicitly include the A:T 
base pairs at the two ends. Therefore, in e.g. 5'-A(CG)$_n$T-3' junctions, we have $N=4n+4$ molecular 
electronic sites within the tight-binding ladder model.

Beyond the molecular electronic structure of the ds-DNA,
three additional parameters should be provided as input to the LBP 
equations:
The position of the Fermi energy $\epsilon_F$ relative to the molecular states,
the strength of metal-molecule hybridization $\gamma_{L,R}$, and an
appropriate value for $\gamma_d$.
In principle, we could use a range of values for $\gamma_d$, 
to capture the susceptibility of different bases 
and sites along the DNA molecule to environmental interactions. 
Here, for simplicity, we use a single value 
for $\gamma_d$, identical for all bases and sites.
While one could carefully optimize these three parameters, $\epsilon_F$, $\gamma_{L,R}$ and $\gamma_d$, 
to reproduce experimental results \cite{Ratner15,Beratan16},
our goal here is to develop a general understanding of transport characteristics as supported by the two  sequences.
Thus, we select reasonable values for these parameters,
and study the transport behavior
with several representative examples. We do not attempt careful 
fitting to experimental results.

%Magnitude, slopes, contrast in the S configurations.

Figure \ref{ladder1} displays LBP calculations for the A and S sequences 
as depicted in  Fig. \ref{schemeA}. %-\ref{schemeS}.
We set  $\epsilon_F=\epsilon_G$,  and use
$\gamma_{L,R}=0.05$ eV and $\gamma_d$=0-0.05 eV.  %XXX EENRGIES
These parameters excellently reproduce 
transport characteristics in the alternating sequence: 
We recover experimental values \cite{Ratner15} for the overall resistance, as well as the resistance per site (slope), 
see panel (a), receiving  $\sim 0.5$ MOhm per unit length for $\gamma_d=0.03$ eV. %DS
Based on the linear increase of resistance with $n$, we argue that 
in alternating sequences charge transport proceed via a site-to-site hopping conduction.
Note that $\gamma_d$ in the range 5-50 meV is comparable to electronic matrix elements in 
DNA structures, see Supporting Information. 
%Coherent and incoherent effects thus compete to support intermediate  transport behavior.
 %
Also, we confirmed that we can reproduce qualitatively
the tunneling-to-hopping crossover in conductance when implanting an (AT)$_m$ block within a GC rich sequence, 
as observed in Refs. \cite{Tao04,Tao16}, see Supporting Information.

Our simulations of stacked sequences in Fig. \ref{ladder1} only  
qualitatively agree with measurements \cite{Ratner15}: 
We observe an even-odd trend in resistance, yet note on a very small slope
with respect to length.
This result should be contrasted by experiments demonstrating 
that S and A sequences support similar resistances per site \cite{Ratner15}.

Considering the interpolating expression Eq. (\ref{eq:RSf}),  we argue that
in alternating sequences the first (hopping) term  dominates, while in S sequences
the second (ballistic) contribution initially controls charge transfer. %MK
In-line with our expectations, stacked structures support coherent-delocalized
conduction, turning into ohmic behavior for long enough chains.

We display in Fig. \ref{eigg} the electronic eigenenergies of A and S sequences with $n=3-7$ units.
Recall that we place the Fermi energy at $\epsilon_F=\epsilon_G=0$, and the energies of the other three bases are set relative to this value.
We find that both A and S sequences support four bands, deriving from the different bases. 
When we zoom-in on the energy levels close to the Fermi energy (lower panels) we 
find that in A sequences the G band separates into two sub-bands, with the 
Fermi energy placed within the gap (order of $k_BT$).
To contrast, in S sequences, the G band (panel s2) does not develop a significant gap.
Furthermore, in stacked structures with $n=3,5,7,...$ the Fermi energy touches a molecular resonance, while
for $n=4,6,...$ the Fermi energy is slightly offset from the Fermi energy.
The energy level structure immediately suggests that in A sequences ohmic-hopping conduction should dominate: charges are injected into the first G site from the metal across a non-zero energy barrier, and
hopping between G sites is thermally activated according to the low barrier ($<0.1$ eV) between G states \cite{Segal}.
In contrast, the S system supports a metallic-like structure for the G band. Charge injected from
the metal is immediately delocalized over the system, and transport through
the molecule is partially ballistic given the metallic band structure, missing gaps. %MK

We repeated the simulations of Fig. \ref{ladder1} while employing parametrization from
Ref. \cite{voityuk} (not shown). In this set, electronic coupling between bases were generated from QM/MD simulations,
by averaging couplings over the MD trajectory.
We found that these parameters \cite{voityuk} generated similar results 
as in Fig. \ref{ladder1}, with the even-odd contrast in stacked sequences moderately amplified.

In Fig. \ref{ladder2} we perform simulations with strong metal-molecule hybridization,
$\gamma_{L,R}$= 1 eV.  In comparison to  Fig. \ref{ladder1}
we find that the resistance of A sequences is almost independent of the contact energy, in support of
the hopping model. In contrast, the resistance of the S sequence is significantly reduced at stronger hybridization.
Also, the even-odd effect almost disappears. 
This is because the ballistic resistance (which is responsible
for the even-odd effect) is very small at large $\gamma$, and the overall
resistance is dominated by the hopping contribution, which is linear with $n$. 

In support of transport calculations, we display in Fig. \ref{DNArdm1} maps of the density matrix in alternating (GC)$_5$
(left) and stacked G$_5$C$_5$ (right) sequences. 
The enumeration begins at the left contact, and we organize the bases as follows, 
$(s=1,j) \rightarrow$ $|2j-1\rangle$, 
$(s=2,j) \rightarrow$ $|2j\rangle$, $j=1,2,...$, with $s$ as the DNA strand (1 is the strand coupled to the left electrode, 
and 2 to the right, see Fig. \ref{schemeA}), and $j$ as the base pair index. 
The color map reveals the following features:
(i) Population is high (red) on the G sites (recall that we set $\epsilon_F=\epsilon_G$).
(ii) In both A and S sequences neighboring G bases maintain their coherences (yellow squares).
(iii) In S sequences, coherences survive even between G's that are placed further apart. 
In fact, delocalization is preserved even {\it across segments}. For example,
 the first G (state $|3\rangle$) maintains its coherence with the last G (state $|22\rangle$), which is placed on the other segment.

We had further analyzed the density matrix of stacked sequences of different lengths 
($n=3-6$). In all cases the density matrix displays extended coherences between G sites. 
We do not however observe clear even-odd signatures (between e.g. $n=5$ and $n=6$), as noted in the 1D model, Fig.  \ref{Figrdm2}.

%More RDM
% Higher gammad.
% check n=3,n=4,n=5.

% cite Peccia paper
%It should be noted that given the 
%Note: no backbone
% Berlin parameters A or B form?
%=====================================
\section{Summary} 
\label{Summ}

We studied in detail the charge transport properties of stacked and alternating G-C DNA sequences
using the Landauer-B\"uttiker probe technique in order to understand the qualitative differences
in their observed transport characteristics. 
The probe was essential in enabling us to model the incoherent hopping
behavior of long DNA chains, which would have been missed by the bare Landauer formalism.
Starting from a simplified 1D Hamiltonian, we were able to construct analytical expressions for the
resistances of different types of near-resonance conductors and gain intuition for the full DNA model.
We further applied our method to more realistic ladder-type DNA Hamiltonians with parameters retrieved from ab-initio simulations.

We found that so-called alternating DNA chains conduct primarily via site-to-site incoherent
hopping, while in stacked sequences there is also an oscillating ballistic contribution, 
owing to the delocalization of molecular orbitals along adjacent G bases. These oscillations
are sensitive to the parity of stacked segments in both the 1D and full-ladder
DNA models. We stress that according to our simulations, this ballistic
transfer, and not superexchange/deep tunneling, constitutes the coherent contribution to 
`intermediate coherent-incoherent' charge transport in  stacked GC sequences. Between 1D and ladder DNA systems,
there was a significant deviation in the agreement between the slopes of resistance curves for A and S molecules. 
This can be immediately explained in terms of the band structure of each system. In the ladder DNA system,
A and S chains form distinctly different band structures near the Fermi energy: The A-type chains form a gap that is greater that $k_BT$ 
while S molecules support a metallic-like band.
%and driving the divergence in the resistance curves.  % DS 'divergence' sounds like a mathematical divergency.
By comparison, the 1D S and A systems form a similar
band structure, which allows them to facilitate hopping currents with comparable resistance per site. %MK

Our results are in qualitative agreement with experiments \cite{Ratner15,Beratan16}, 
though our simulations of transport in stacked sequences underestimate the electrical resistance per site.
%This should not be a surprise, since given the band structure in Fig. \ref{eigg}, which is essentially metallic,
%on-resonance tunneling out-compete hopping current for values of $\gamma_d$ within the band.
% DS
%(at any realistic $\gamma_d$).  %MK - Dvira double-check these statements & MK check the incoherent ratio
The observed deviations from experiment could result from several simplifying assumptions:
(i)  We assumed that the electronic structures of the two families, A and S, are identical and static. 
However, studies show that these sequences may organize differently:
with `structure A' and `structure B' for stacked and alternating sequences, respectively \cite{ABform1, ABform2}.
A more accurate estimate of the resistance properties may then be retrieved by first generating 
electronic parameters specific to each sequence. Also, one should modify $\gamma_d$ between A and S 
sequences, to capture the specific characteristics (i.e. flexibility) of each structure.
(ii) We assumed that environmental effects are uncorrelated and act
identically on all bases. This assumption
could be addressed by extracting $\gamma_d^{k}$, for each base $k$, 
from quantum mechanics/molecular mechanics simulations, 
to characterize electronic (spatial and temporal) fluctuations by simple analytic correlation functions  
\cite{Beratan16}.

In our simulations of double-stranded DNA using a ladder Hamiltonian, we assigned the parameters
$\gamma_d$, $\gamma_{L,R}$ and $\epsilon_F$ so as to generate robust and physically meaningful results. 
The Fermi energy 
$\epsilon_F$ was taken to be near  the site energy of  G bases, which are believed to constitute the primary 
charge transfer pathway for both hopping and ballistic current.
This was also an important choice, since as we learned through study of the 1D model,
ICI behavior with even-odd effects is primarily found close to molecular resonances.
$\gamma_d$ and $\gamma_{L,R}$ were taken within the range of values comparable to molecular parameters.
$\gamma_d$ in particular is of the same order as the G-G electronic coupling for 
stacked sequences (see Supporting Information), which allows incoherent hopping and coherent even-odd effects
to co-mingle, providing insight into ICI behavior. 
%Further, $\gamma_d$ is also of an order of the site energy correlation time as computed for stacked
%sequences via first-principles simulations \cite{Beratan16,Gutierrez1,Kubar1}.
Our results were also quite robust to fluctuations of $\gamma_{L,R}$ within reasonable limits.

We emphasize that unlike quantum mechanics/molecular mechanics  \cite{cunibertiJCP09,cunibertiNJP10}
and stochastic Schr\"odinger equation-based  \cite{Beratan16} calculations, the LBP method employed here
extends beyond the coherent limit, incorporating environmental/incoherent effects within 
the probe transport formalism.
The LBP technique is advantageous here for its significant simplicity,
including essentially a single 'fitting parameter' ($\gamma_d$) and incorporating 
incoherent effects in a straightforward way. 
%The probe technique has also been shown to phenomenologically reproduce many key features of molecular conduction
%in different transport regimes. %MK probe vs. other methods - think about whether to keep
In the future we will explore the onset of the intermediate 
coherent-incoherent regime in other DNA strands, 
e.g., explore chiral effects by swapping the 3'5' orientation with respect to the leads, 
and by exploring the role of mismatches and  disorder.
It is interesting as well to revisit the problem of long-range (almost distance independent) charge 
transfer in certain DNA sequences \cite{GieseE}, and employ the probe method to gain 
further insights on the role of the environment in highly efficient charge propagation processes 
\cite{PeskinUn}.

%===================================
\begin{acknowledgments}
The work was supported by the Natural Sciences and Engineering Research Council of Canada and the Canada Research Chair Program.
The work of Michael Kilgour was partially funded by an Ontario Graduate Scholarship. 
Hyehwang Kim was supported by an
Ontario Student Opportunity Trust Funds Research Scholarship.
The authors thank Mark A. Ratner, Rafael Gutierrez, Uri Peskin, and Theodore Zwang for helpful discussions.
\end{acknowledgments}

%============================
{}

\newpage

%\begin{widetext}
\setcounter{equation}{0}
\setcounter{section}{0}
\setcounter{figure}{0}
\clearpage
{\bf Supporting Information for ``Intermediate coherent-incoherent charge transport: DNA as a case study"}

%\begin{document}

%\maketitle

\renewcommand{\thepage}{S\arabic{page}} 
\renewcommand{\thesection}{S\arabic{section}}  
\renewcommand{\thetable}{S\arabic{table}}  
\renewcommand{\thefigure}{S\arabic{figure}} 
\renewcommand*{\citenumfont}[1]{S#1}
\renewcommand*{\bibnumfmt}[1]{[S#1]}
%======================================================================
\section{Calculations with shifted Fermi energy}

We simulate here the resistances of the A and S sequences while
shifting the position of the Fermi energy away from the bridge energy $\epsilon_B$ (1D model) and the energy of the
G base $\epsilon_G$ (ladder model).

\begin{figure*}[ht]
\vspace{0mm} \hspace{-6mm}
{\hbox{\epsfxsize=170mm \epsffile{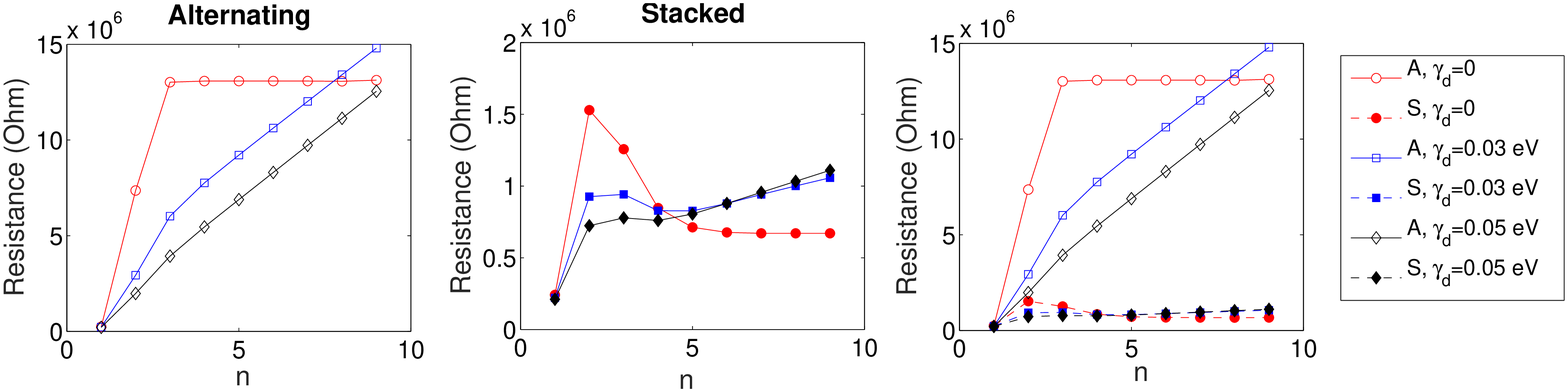}}} %g3Epn35.eps
\caption{1D model.
Resistance as a function of length for (a) alternating and (b) stacked sequences.
Panel (c) overlays A and S results.
We used $\epsilon_F-\epsilon_B=0.35$, $v_s=0.14$, $v_a=0.1$, $\gamma_{L,R}=0.3$ all in eV, room temperature.
}
\label{FigA1}
\end{figure*}

Concerning the 1D model described in Sec. 2,  we repeat the calculation of Fig.
2 (where we set $\epsilon_B=\epsilon_F$) and study in Fig. \ref{FigA1} the case with
$\epsilon_F-\epsilon_B=0.35$ eV.
We note that, as expected, the resistances in off-resonance structures are higher than those
obtained in on-resonance situations, for both A and S structures.
Specifically, for $\gamma_d=0$ the A sequence demonstrates a strong enhancement of resistance with size for 
short systems  (corresponding to tunneling conduction), and a ballistic behavior beyond $n=2$. 
At finite $\gamma_d$,  ohmic behavior dominates beyond $n=2$.
The S system displays a rather nontrivial behavior with a striking nonmonotonic  trend;
the resistance rises sharply for $n=1-2$, and it {\it decreases} for $n=3-4$,
beyond which it grows linearly with size. %in accord with the hopping conduction mechanism.
It is also significant to note that S-type molecules show a significantly higher conductance values relative to A-type molecules.
All in all, it is interesting to note that S sequences do not display a typical tunneling-to-hopping 
crossover even when $|\epsilon_B-\epsilon_F|>k_BT,v_a,v_s$.

%=============================
% figure

%==============================
% figure
\begin{figure*}[htbp]
\vspace{0mm} \hspace{-6mm}
{\hbox{\epsfxsize=185mm \epsffile{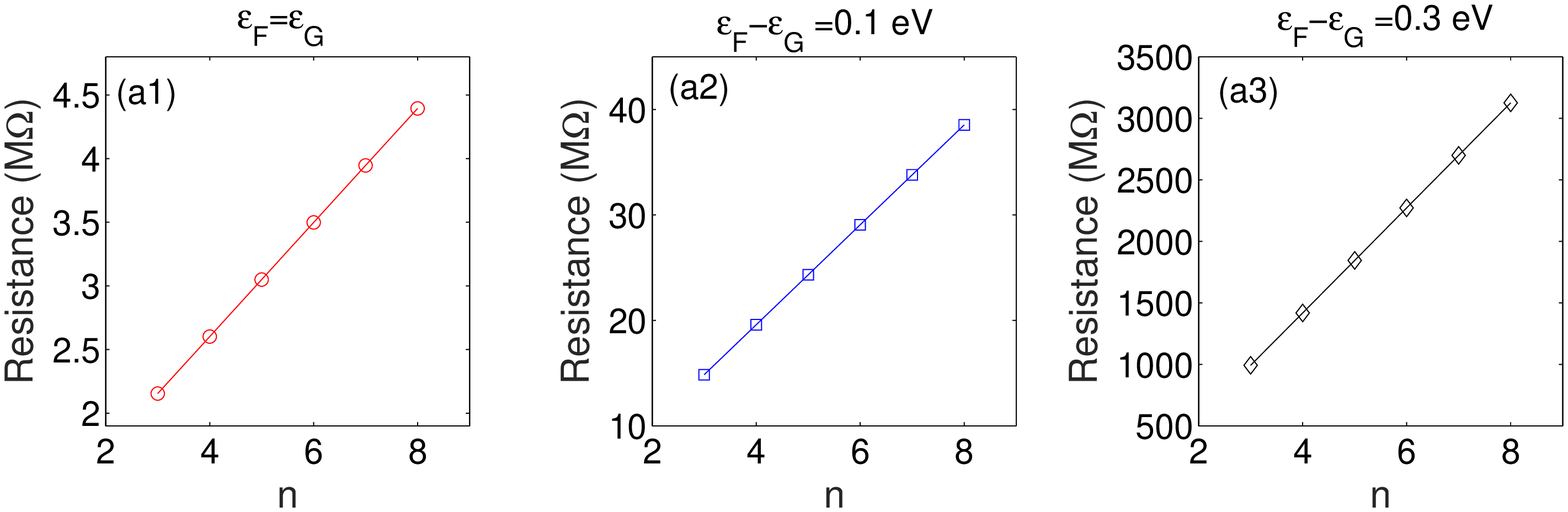}} %ErefA1.eps
\hbox{\epsfxsize=185mm \epsffile{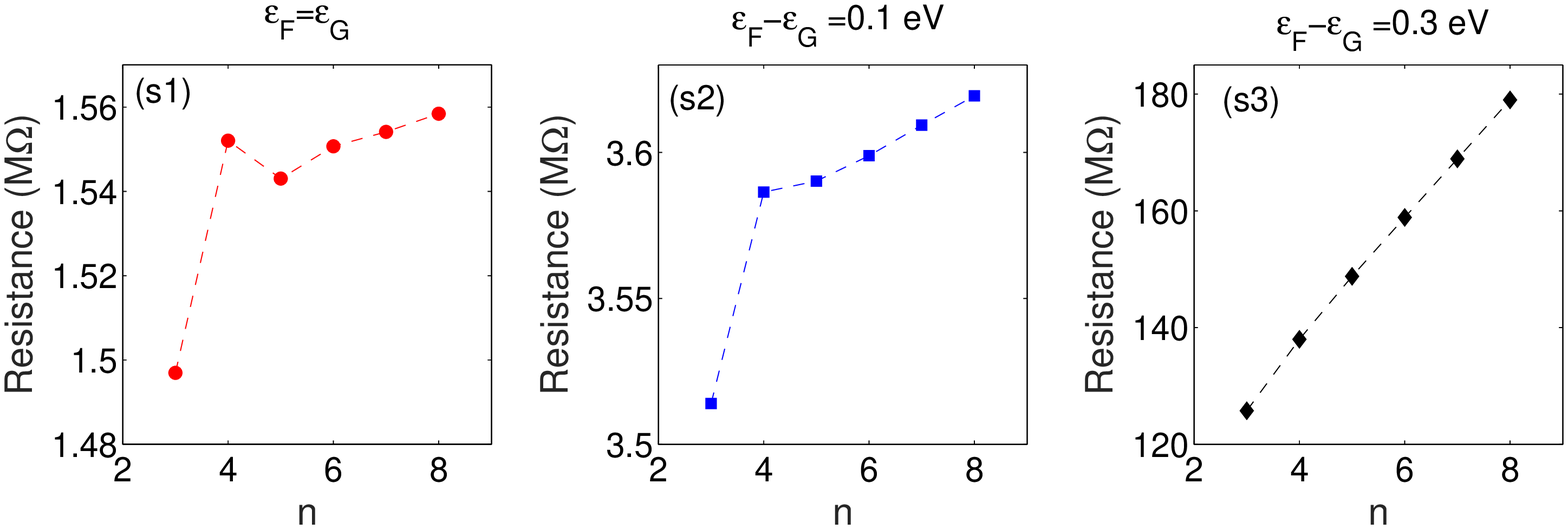}}}
\caption{Ladder model.
Resistance as a function of length for the alternating (a1)-(a3) and stacked (s1)-(s3) sequences
with $\epsilon_F-\epsilon_G=0$,
$\epsilon_F-\epsilon_G= 0.1$ eV, $\epsilon_F-\epsilon_G=0.3$ eV,
as identified in the panels.
Other parameters are $\gamma_d=0.03$ eV and $\gamma_{L,R}=0.05$ eV, room temperature.
}
\label{FigA2}
\end{figure*}

%=========================================

Turning our attention to the  ladder model of Sec. 4, we recall that in Figs. 8-9
we set the Fermi energy within the G band, $\epsilon_F=\epsilon_G$.
We further identified in Fig. 10 the position of the Fermi energy with respect to the electronic eigenenergies.
In Fig. \ref{FigA2} we examine the resistance of A and S sequences when $\epsilon_F-\epsilon_G\neq 0$.
We use $\gamma_d= 0.03$ eV, which quantitatively reproduces experimental
results for the A sequence \cite{Ratner15}. 
We find that a shift of the Fermi energy by 0.1 eV away from the G band 
results in a significant enhancement of resistance for both A and S sequences.
For the S sequence in particular we note that by gradually shifting the Fermi energy from the G band,
the even-odd oscillations fade out, fully disappearing when $\epsilon_F-\epsilon_G=0.3$ eV.

It was demonstrated in Fig. 10 that both A and S sequences support four energy bands. 
However, for the A sequence the G band separates into two sub-bands with the Fermi energy positioned within the gap.
For S sequences there is no gap in the G band. 
By shifting the  Fermi energy away from the G band---such that it now lies between the first two energy bands---the 
charge injected from the metal is no longer delocalized over the system. As a result,
the two structures A and S similarly conduct---following the ohmic-hopping mechanism. 
This behavior is illustrated in Fig. \ref{FigA2}: When $\epsilon_F-\epsilon_G$ = 0.3 eV, the
 resistance increase linearly with length for both S and A DNA molecules.

Based on these simulations we conclude that in relevant experiments \cite{Ratner15,Beratan16} the molecular electronic states
participating in transport lie close to the Fermi energy. 
This is supported by two observations: (i) When $\epsilon_G=\epsilon_F$ we receive
resistances for A sequences in a good agreement with experiments. (ii) In S sequences
the ballistic contribution survives on-resonance, but once we place $\epsilon_F-\epsilon_G\gtrsim 0.1$ eV,
even-odd oscillations disappear and the resistance follows a monotonic linear-Ohmic trend.

%==============================

% figure
%\begin{center}
%\begin{figure*}[hbp]
%\vspace{0mm} \hspace{3mm}
%{\hbox{\epsfxsize=185mm \epsffile{ErefS.eps}}} %ErefS1.eps
%\caption{
%Resistance as a function of length for the stacked (S) sequence with (a) $\epsilon_F-\epsilon_G=0$,
%(b) $\epsilon_F-\epsilon_G=-0.1 eV$, (c) $\epsilon_F-\epsilon_G=-0.3 eV$.
%We used $\gamma_d=0.03$, and $\gamma_{L,R}=0.05$ all in eV, and room temperature.
%}
%\label{FigA3}
%\end{figure*}
%\end{center}
%===========================================

%==============================

\section{Electronic matrix elements for the ladder model}

For the completeness of our presentation we compile three tables  \ref{table:1}-\ref{table:3} from Ref. \cite{BerlinJacs}
with values  for charge-transfer integrals as used in our simulations for the ladder model of double-stranded DNA.
As noted in the literature, $t_{5'-XY-5'}=t_{5'-YX-5'}$ and $t_{3'-XY-3'}=t_{3'-YX-3'}$ 
from symmetry, but $t_{5'-XY-3'} {\neq} t_{3'-XY-5'}$, given the  directionality of DNA \cite{BerlinJacs}.

%================================================================================================
\vspace{5mm}
\begin{table}[h!]
\begin{tabular}{c c c c c} 
\hline
\hline
&& \hspace{8mm}{\bf Y}&&\\
\hline
{\bf X} \hspace{1.4mm} \vline & G & A & C  & T\\ [0.5ex] 
\hline
G \hspace{1.8mm} \vline & 0.053 & -0.077 & -0.114 & 0.141 \\ 
A \hspace{2mm} \vline & -0.010 & -0.004 & 0.042 & -0.063\\
C \hspace{2mm} \vline & 0.009 & -0.002 & 0.022 & -0.055\\
T \hspace{2mm} \vline & 0.018 & -0.031 & -0.028 & 0.072\\ [1ex] 
\hline
\hline
\end{tabular}
\caption{$t_{5'-XY-3'}=t_{3'-YX-5'}$ $(eV)$  \cite{BerlinJacs}}
\label{table:1}
\end{table}

%=========================================================================================================
\vspace{5mm}
\begin{table}[h!]
\begin{tabular}{c c c c c} 
\hline
\hline
&& \hspace{8mm}{\bf Y}&&\\
\hline
{\bf X} \hspace{1.4mm} \vline & G & A & C  & T\\ [0.5ex] 
\hline
G \hspace{1.8mm} \vline & 0.012 & -0.013 & 0.002 & -0.009 \\ 
A \hspace{2mm} \vline & -0.013 & 0.031 & -0.001 & 0.007\\
C \hspace{2mm} \vline & 0.002 & -0.001 & 0.001 & 0.0003\\
T \hspace{2mm} \vline & -0.009 & 0.007 & 0.0003 & 0.001\\ [1ex] 
\hline
\hline
\end{tabular}
\caption{$t_{5'-XY-5'}$ $(eV)$  \cite{BerlinJacs}}
\label{table:2}
\end{table}

%=========================================================================================================

\begin{table}[h!]
\begin{tabular}{c c c c c} 
\hline
\hline
&& \hspace{8mm}{\bf Y}&&\\
\hline
{\bf X} \hspace{1.4mm} \vline & G & A & C  & T\\ [0.5ex] 
\hline
G \hspace{1.8mm} \vline & -0.032 & -0.011 & 0.022 & -0.014 \\ 
A \hspace{2mm} \vline & -0.011 & 0.049 & 0.017 & -0.007\\
C \hspace{2mm} \vline & 0.022 & 0.017 & 0.010 & 0.004\\
T \hspace{2mm} \vline & -0.014 & -0.007 & 0.004 & 0.006\\ [1ex] 
\hline
\hline
\end{tabular}
\caption{$t_{3'-XY-3'}$ $(eV)$  \cite{BerlinJacs}}
\label{table:3}
\end{table}

%=========================================================================================================
%Is this method better? I can't seem to make this table look nice.

%\begin{table}[h!]
%\begin{tabular}{c c c c c} 
%\hline
%$t_{5'-XY-3'}=t_{3'-YX-5'}$ $(eV)$ \\
%\hline
%\hspace{20mm}{\bf Y}&\\
%\hline
%{\bf X} \hspace{1.4mm} \vline & G & A & C  & T\\ [0.5ex] 
%\hline
%G \hspace{1.8mm} \vline & 0.053 & -0.077 & -0.114 & 0.141 \\ 
%A \hspace{2mm} \vline & -0.01 & -0.004 & 0.042 & -0.063\\
%C \hspace{2mm} \vline & 0.009 & -0.002 & 0.022 & -0.055\\
%T \hspace{2mm} \vline & 0.018 & -0.031 & -0.028 & 0.072\\ [1ex] 
%\hline
%$t_{5'-XY-5'}$ $(eV)$ \\
%\hline
%G \hspace{1.8mm} \vline & 0.012 & -0.013 & 0.002 & -0.009 \\ 
%A \hspace{2mm} \vline & -0.013 & 0.031 & -0.001 & 0.007\\
%C \hspace{2mm} \vline & 0.002 & -0.001 & 0.001 & 0.0003\\
%T \hspace{2mm} \vline & -0.009 & 0.007 & 0.0003 & 0.001\\ [1ex] 
%\hline
%$t_{3'-XY-3'}$ $(eV)$ \\
%\hline
%G \hspace{1.8mm} \vline & -0.032 & -0.011 & 0.022 & -0.014 \\ 
%A \hspace{2mm} \vline & -0.011 & 0.049 & 0.017 & -0.007\\
%C \hspace{2mm} \vline & 0.022 & 0.017 & 0.010 & 0.004\\
%T \hspace{2mm} \vline & -0.014 & -0.007 & 0.004 & 0.006\\ [1ex] 
%\hline
%\end{tabular}
%\caption{  \cite{BerlinJacs,Wolf}}
%\label{table:1}
%\end{table}

%====================================================================================================

\section{Simulations of tunneling-to-hopping crossover in DNA}

We support our LBP approach and the DNA parametrization by further simulating
the tunneling-to-hopping conductance crossover in other DNA sequences.
Measurements show that when placing (AT)$_m$ units at the center of GC sequences,
the resistance follows a tunneling (superexchange) behavior---for short AT blocks---since
 AT bases act as a tunneling barrier \cite{Tao16}. For longer blocks of $m=3-4$ units,
hopping-Ohmic resistance dominates the transport behavior.

\begin{center}
\begin{figure*}[ht]
\vspace{0mm} \hspace{0mm}
{\hbox{\epsfxsize=130mm \epsffile{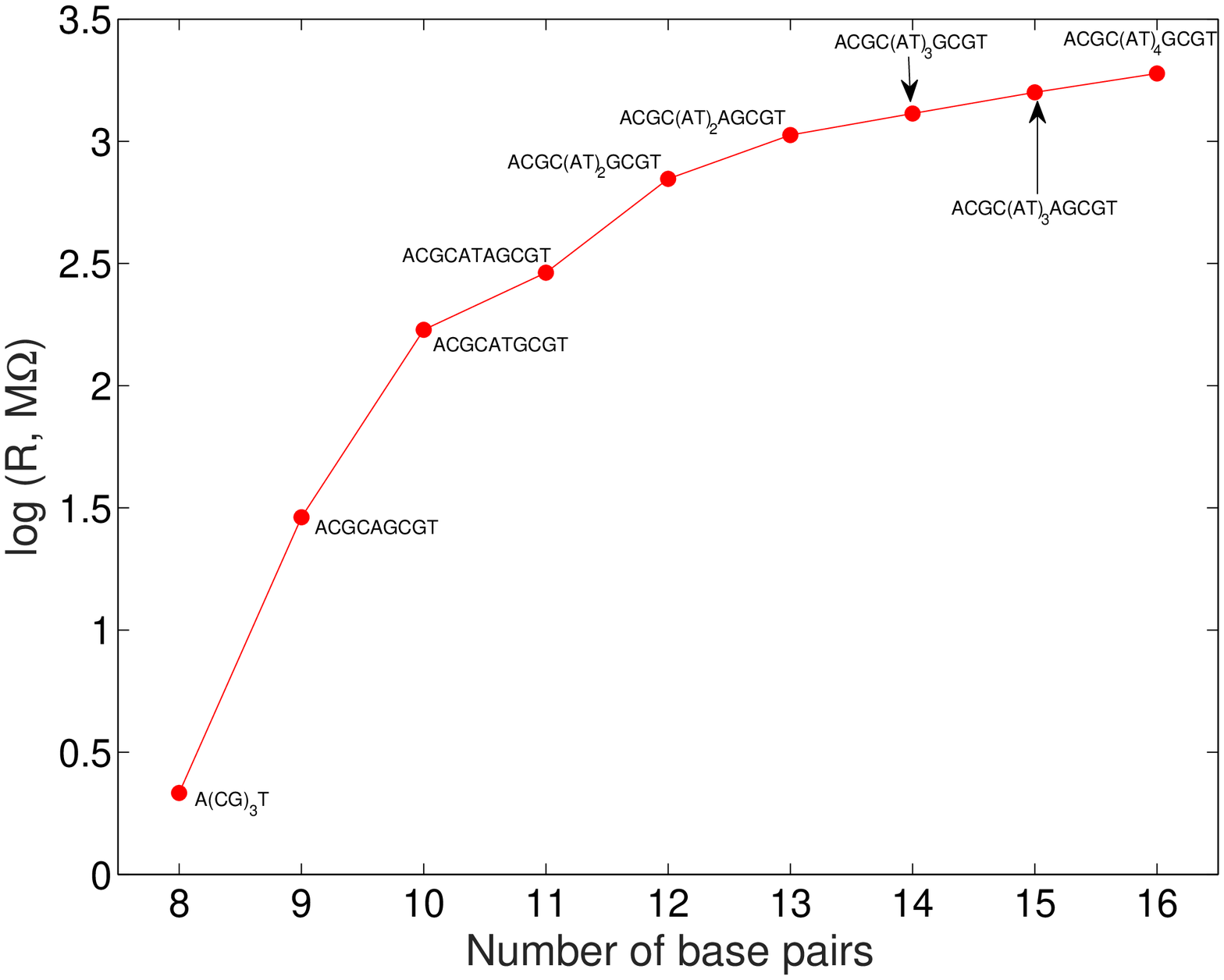}}} %ErefS1.eps
\caption{
Resistance vs. the number of base pairs in sequences with AT blocks of different sizes, inserted into a GC domain.
We use $\gamma_{L,R}=0.05$ eV and $\gamma_{d}=0.03$ eV, $\epsilon_F=\epsilon_G$, room temperature.
}
\label{FigA3}
\end{figure*}
\end{center}

We use parameters from Table 1 (main text) and Tables \ref{table:1}-\ref{table:3}, 
and study the resistance of
several sequences, as identified in Fig. \ref{FigA3}. We employ the LBP method with 
$\gamma_d$=0.03 eV and $\gamma_{L,R}=0.05 $ eV, parameters used to receive the resistance of alternating sequences in Fig. 8.
Our simulations demonstrate a sharp increase in resistance with length, for short molecules, followed by a more moderate enhancement
in longer systems, in accord 
with the tunneling-to-hopping crossover behavior reported in Ref. \cite{Tao16}.
However, we note that our simulated resistances are significantly larger than experimental values \cite{Tao16}. 
We emphasize though that we did not optimize our LBP results through a careful tuning of $\epsilon_F$, $\gamma_d$ and $\gamma_{L,R}$.

% figure

%====================================================================================================

%\end{widetext}
%====================================

\end{document}